\documentclass[aps,prl,twocolumn,superscriptaddress,showpacs]{revtex4}
\usepackage{graphicx}
\usepackage{latexsym}
\usepackage{amssymb}
\usepackage{amsmath}
\usepackage{amsfonts}
\usepackage{bm}
\usepackage{multirow}
\usepackage{color}
\usepackage{xcolor}
\usepackage{tikz}
\usepackage{tikz-feynman}

\usepackage[colorlinks=true, citecolor={blue!80!black}, urlcolor={blue!50!black}, linkcolor = {blue!80!black}]{hyperref}

\usepackage[percent]{overpic}
\usepackage{tabularx}

\newcommand{\be}{\begin{equation}}
\newcommand{\ee}{\end{equation}}
\newcommand{\bea}{\begin{eqnarray}}
\newcommand{\eea}{\end{eqnarray}}
\newcommand{\pd}{{\vphantom\dagger}}

\begin{document}

\title{Dynamical signatures of quasiparticle interactions in quantum
  spin chains}

\author{Anna Keselman}
\affiliation{Kavli Institute for Theoretical Physics, University of
  California, Santa Barbara, CA 93106, USA}
\author{Leon Balents}
\affiliation{Kavli Institute for Theoretical Physics, University of
  California, Santa Barbara, CA 93106, USA}
\affiliation{Canadian Institute for Advanced Research, Toronto,
  Ontario, Canada}
\author{ Oleg A. Starykh}
\affiliation{Department of Physics and Astronomy, University of Utah,
  Salt Lake City, Utah 84112, USA}

\begin{abstract}
  We study the transverse dynamical susceptibility of an
  antiferromagnetic spin$-1/2$ chain in presence of a longitudinal
  Zeeman field.  In the low magnetization regime in the gapless phase,
  we show that the marginally irrelevant backscattering interaction
  between the spinons creates a non-zero gap between two branches of
  excitations at small momentum. We further demonstrate how this gap
  varies upon introducing a second neighbor antiferromagnetic
  interaction, vanishing in the limit of a non-interacting ``spinon
  gas".  In the high magnetization regime, as the Zeeman field
  approaches the saturation value, we uncover the appearance of
  two-magnon bound states in the transverse susceptibility. This
  bound state feature generalizes the one arising from string states
  in the Bethe ansatz solution of the
  integrable case.  Our results are based on
  numerically accurate, unbiased matrix-product-state techniques as
  well as analytic approximations.
\end{abstract}

\maketitle

\emph{Introduction-} The nearest-neighbor antiferromagnetic spin-1/2
chain~\cite{Bethe1931} serves as a paradigmatic model of strongly
correlated systems.  In particular, it is an established setting in
which the existence of {\em fractionalized} quasiparticles, deemed
spinons, is incontrovertible.  Spinons are the elementary excitations
for magnetic fields below the critical saturation field, above which
the ground state becomes fully polarized.  In the saturated regime,
the elementary excitations are instead simple (not fractionalized)
magnons, or spin waves.  A quasiparticle is, by definition, an
excitation which, when isolated, propagates freely with minimal decay.
However, in general, when multiple quasiparticles are present, they
interact, and in a strongly correlated system, they interact strongly.
Such interactions are present both for fractionalized and ordinary
excitations, but impact the physics particularly strongly in the
former case, as any local operator creates in this case more than one
quasiparticle.  

In this manuscript, we focus on dynamical signatures,
i.e. features in the frequency-dependent spin correlations, of such
quasiparticle interactions, in presence of a magnetic field.  
Extensive numerical
studies~\cite{Muller1981,Kohno2009,Yang2019,Nishimoto2007,Bouillot2011,Lefmann1996}
of the dynamical correlation functions have been carried out in the
Heisenberg limit, for which a Bethe ansatz solution exists.
Consequently the general structure of the correlations, and their
deconstruction in terms of spinons, is already well established.  
Here, we go beyond the integable limit, and add to this broad picture a detailed analytical theory of the effects of quasiparticle interactions, in both the low and high
magnetic field (magnetization) regimes.
Specifically, in the low
magnetization case, we employ a continuum field theory in which spinon
interactions are parametrized by the marginally irrelevant
current-current backscattering coupling $g$.  We show that this
coupling creates a gap between two branches of excitations at zero
momentum, which is equal to $g M$, where
$M$ is the magnetization, thus directly revealing the strength of interactions between spinons.
We note that the fermionic field theory used here and the
results we obtain for it have a direct parallel to a recent study by
two of us on two dimensional spin liquids with a spinon Fermi
surface~\cite{BalentsStarykh2020}. The results herein thus serve in part as a proof of
principle for the two dimensional case, with the major advantage that here
the results are vetted by unbiased numerical simulations.

In the regime of large magnetization,
the natural quasiparticles are spin flip magnons (spins
anti-aligned with the field), leading to a dominant single branch in
the dynamical susceptibility.  We show however that two-magnon bound
states appear as an additional higher energy branch due to interaction
of the ``probe'' magnon with those already present in the ground
state.  The spectral weight of this higher energy branch is thus
directly proportional to the strength of interactions, as well as the
density of ground state magnons.  Previous studies of
the Heisenberg chain~\cite{Muller1981,Kohno2009} have indeed
identified the Bethe ansatz string solutions to be related to
two-magnon bound states in the high magnetization regime. However,
here we show that this is not unique to the integrable chain limit,
and provide a simple understanding which does not require the
specialized Bethe ansatz formalism.

We test these analytical predictions by comparison to computations using numerical
matrix-product-state (MPS) techniques~\cite{Schollwoeck2011}.  In the small magnetization regime, we tune
the spinon interaction $g$ by including a second-neighbor exchange
coupling $J_2$, while in the large magnetization regime, magnon-magnon interactions are tuned
by introducing magnetic anisotropy of the XXZ form.
In both limits MPS calculations compare excellently with
the theoretical predictions as the respective parameters controlling the density of quasiparticles and the strength of interactions between them are varied.

\emph{Model --}
We consider a spin-$1/2$ chain, with antiferromagnetic nearest-neighbor coupling, $J_1>0$, and next-nearest-neighbor coupling, $J_2$, in longitudinal Zeeman field, $B$. The Hamiltonian of the system is given by
\begin{equation}\label{eq:H}
H=\sum_{i} J_1 \left(\vec{S}_i\cdot \vec{S}_{i+1}\right)_{\eta} + J_2 \left( \vec{S}_i\cdot \vec{S}_{i+2} \right)_{\eta} - B S^z_i,
\end{equation}
where $\vec{S}_i$ is a spin-$1/2$ operator on site $i$.
We allow for anisotropic interactions and denote $\left(\vec{S}_i\cdot \vec{S}_j\right)_{\eta} \equiv S^x_i S^x_j+S^y_i S^y_j+\eta S^z_i S^z_j$.
In the isotropic case, $\eta=1$, and for $B=0$, the system undergoes
a phase transition at $J_2=J_{2,c}\approx 0.241J_1$, between a gapless
and a dimerized phase~\cite{Okamoto1992,Eggert1996}. In the following
we will consider the regime $J_2\leq J_{2,c}$ in which the system remains gapless.

We study the transverse susceptibility $\chi^\pm(k,\omega)$ imaginary part of which determines dynamical structure factor 
$S^{+-}(k,\omega)$ of the dynamical correlations at zero temperature, namely
\begin{eqnarray}
\label{eq:Spm}
S^{+-}(k, \omega) & = & \int_{-\infty}^{\infty} d t e^{i \omega t} \sum_{j=-\infty}^{\infty} e^{-i k j}\left\langle S^+_{j}(t) \cdot S^{-}_{0}(0)\right\rangle \nonumber \\
 & = & \sum_m \left| \left\langle m\right| S^{-}_k \left|0\right\rangle \right|^2\delta(\omega-E_m),
\end{eqnarray}
where  $\left|0\right\rangle$ denotes the ground state of the system.

Our numerical calculations are carried out using the ITensor library~\cite{ITensor}.
To obtain the spectral function~\eqref{eq:Spm} we first obtain the ground state of the system using density matrix renormalization group (DMRG)~\cite{White1992}. We then perform time evolution up to times $t_{\rm max}=80 J_1^{-1}$ using time evolving block decimation (TEBD)~\cite{Vidal2004}. Our analysis is done on finite systems of length $N=400$ sites with open boundary conditions (see SM for further details).

\emph{Low magnetization --}
In the discussion below we focus on the isotropic
case, i.e. $\eta=1$.  The low energy effective description of the
$J_1-J_2$ chain is given by an $SU(2)_1$ Wess-Zumino-Witten conformal
field theory.  As discussed in Ref.\cite{gogolin2004bos,starykh-furusaki-balents,maslov2015}, 
this theory has a convenient fermion representation which is faithful
and simple for the Hamiltonian and spin currents we study here.  We
denote the right/left moving fermionic spinons which constitute the
low energy theory by $\psi_{R/L,s}$, where $s=\uparrow,\downarrow$ is
the spin. The respective spin current is given by
$\vec{J}_{R} = \frac{1}{2} \psi_{R}^\dagger \vec{\sigma}
\psi_R^{\vphantom\dagger}$, where $\psi_R$ denotes two-component
spinor $\psi_R = (\psi_{R\uparrow}, \psi_{R\downarrow})^{\rm T}$ (and
similarly for $\psi_L$).  The low energy Hamiltonian is given by
$H = H_0 + V$, where $H_0$ corresponds to the non-interacting part
\begin{equation}
H_0 = v \int dx \, [\psi_R^\dagger (-i \partial_x) \psi_R^{\vphantom\dagger} + \psi_L^\dagger (i \partial_x) \psi_L^{\vphantom\dagger}]
\end{equation}
(here $v$ is the Fermi velocity), and $V$ is the backscattering interaction
\begin{equation}
V = %-g \int \!dx\, \vec{J}_R\cdot\vec{J}_L = 
- g \int \!dx\,\left[
  J_R^z J_L^z +  \tfrac{1}{2}J_R^+J_L^- +\tfrac{1}{2} J_R^-J_L^+\right].
  \label{eq:1}
\end{equation}
The Hamiltonian $H_0+V$ appears as an interacting fermion problem for
the spinons, an approach we will follow below.  In a standard
bosonization framework $g$ gives rise to a non-linear cosine term. 
In a renormalization group (RG)
treatment, $g>0$ is marginally irrelevant and as a function of its energy argument $E$ flows logarithmically to zero at
low energies \cite{gogolin2004bos,affleckoshikawa1999,OshikawaAffleck2002}, producing
subtle logarithmic modifications to the
temperature dependence of
thermodynamic quantities such as susceptibility and specific
heat~\cite{PhysRevLett.84.4701,PhysRevLett.73.332,lukyanov1998low,starykh1997,takigawa1997}.   
The bare value of $g$ depends on $J_2$ and changes sign at the
critical value, i.e. $g\sim c(J_{2,c}-J_2)$ with $c>0$~\cite{Eggert1996}.

\begin{figure}[t]
	\begin{overpic}[width=0.51\columnwidth]{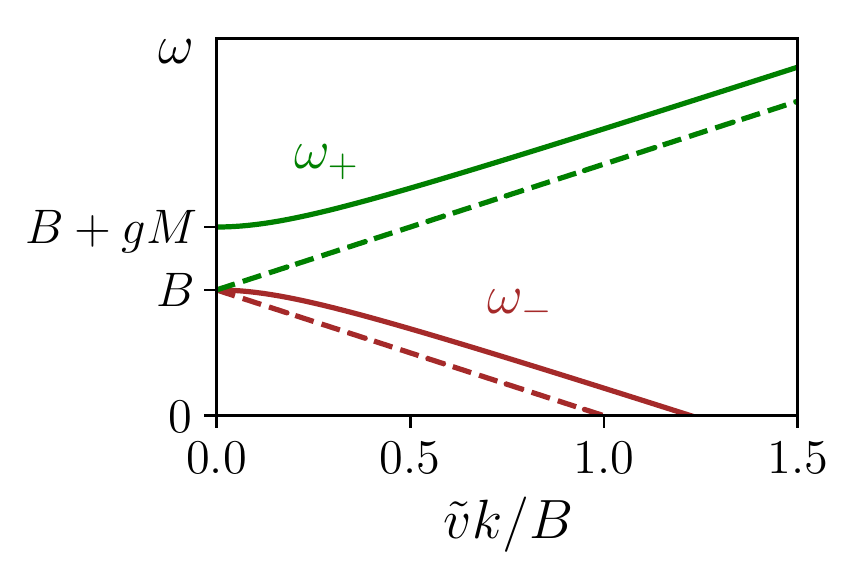} \put (0,60) {\footnotesize{(a)}} \end{overpic}
	\begin{overpic}[width=0.473\columnwidth]{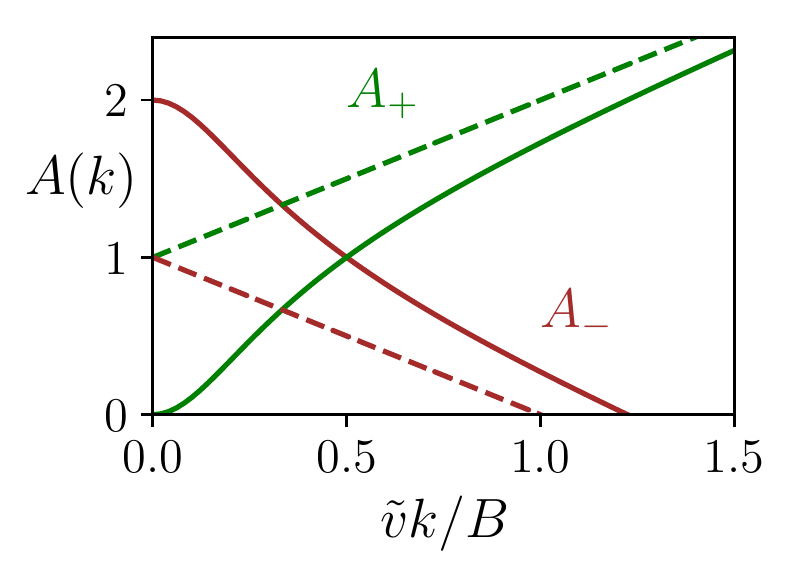} \put (0,65) {\footnotesize{(b)}} \end{overpic}
    \caption{Transverse susceptibility $\chi^{\pm}(k,\omega)$ obtained in the small $k$, and low magnetization regime. (a) The dispersion $\omega_{\pm}(k)$ is given by the green (brown) solid line for $gM/B=1/2$ and green (brown)  dashed line for $g=0$. (b) The intensity of the upper (lower) branch $A_{\pm}$ is the green (brown)  solid line for $gM/B=1/2$ and green (brown)  dashed line for $g=0$. }
     \label{fig:chi_pm_schematic}
\end{figure}

As we now show, the consequences of the non-zero $g>0$ are more
dramatic and directly evident in the spectral features in the presence
of a Zeeman field.  A longitudinal Zeeman field couples to the
magnetization $M$, which is the sum of the right and left spin
currents
\begin{equation}
H_B = - B \int dx  M,\qquad M = J^z_R + J^z_L.
\end{equation}
In the presence of the Zeeman field, RG flow of $g$ toward zero is cut off at $E=B$ \cite{giamarchi2003,OshikawaAffleck2002,SM} and moreover 
distinguishes the effects of the diagonal and spin
flip components of the interactions in Eq.~\eqref{eq:1}. 
Consequently,  we must consider them separately and carefully.  First
let us take $g=0$ and introduce the Zeeman field $B$.  In the spinon
framework, the Zeeman field simply induces a spin splitting of the two
spinon bands.  The dynamical susceptibility is then 
\begin{equation}
\label{eq:chi_pm_g0}
\chi^\pm_0(k,\omega) = \frac{M + \chi_0 v k}{\omega - B - v k}  + \frac{M - \chi_0 v k}{\omega - B + v k} \underset{k\to 0}{\longrightarrow} \frac{2M}{\omega - B},
\end{equation}
where $\chi_0 = 1/(2\pi v)$ is the non-interacting uniform susceptibility.
There
are two branches with linear dispersion and constant spectral weight,
as shown by the dashed lines in Fig.~\ref{fig:chi_pm_schematic}.  Note
that the form at $k=0$ is more robust than for $k>0$, and is in fact
exact provided the Hamiltonian in the absence of Zeeman field has
SU(2) symmetry.  This ``Larmor theorem''\cite{OshikawaAffleck2002}
follows simply from the fact that in this case $[S^+_{\rm tot},H] = B
S^+_{\rm tot}$, where $S^+_{\rm tot}=\sum_i S_i^+$.

\begin{figure}
    \centering
    \begin{overpic}[width=0.48\columnwidth]{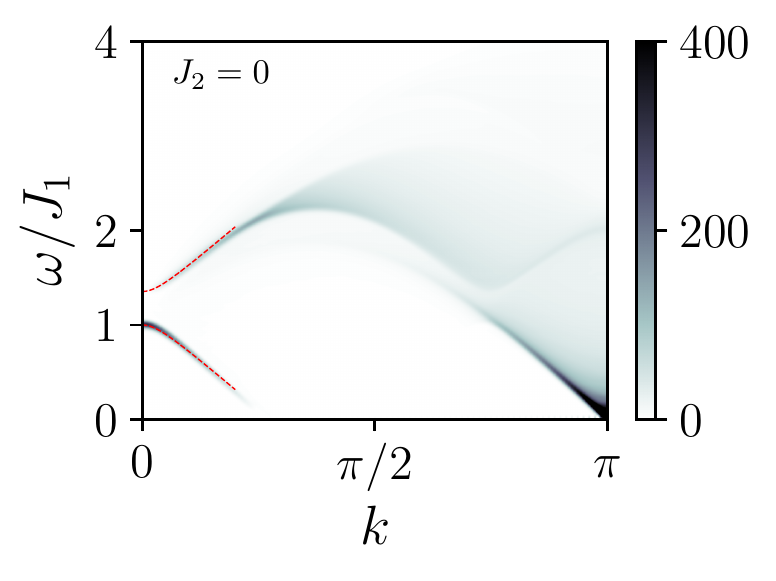} \put (0,70) {\footnotesize{(a)}} \end{overpic}
    \begin{overpic}[width=0.5\columnwidth]{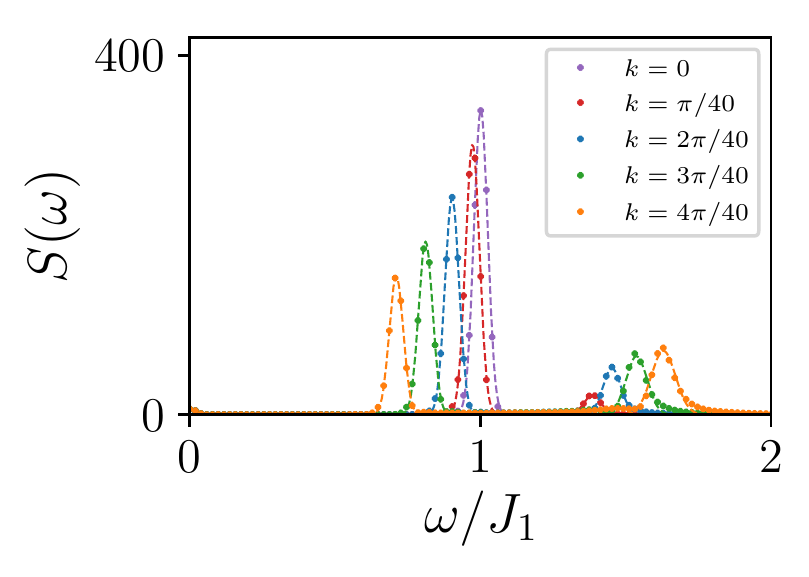} \put (0,67) {\footnotesize{(b)}} \end{overpic} \\
    \begin{overpic}[width=0.48\columnwidth]{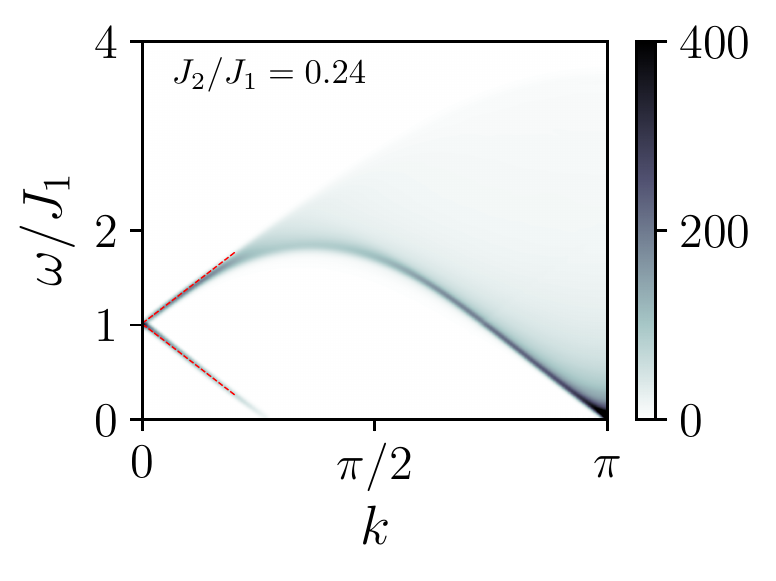} \put (0,70) {\footnotesize{(c)}} \end{overpic}
    \begin{overpic}[width=0.5\columnwidth]{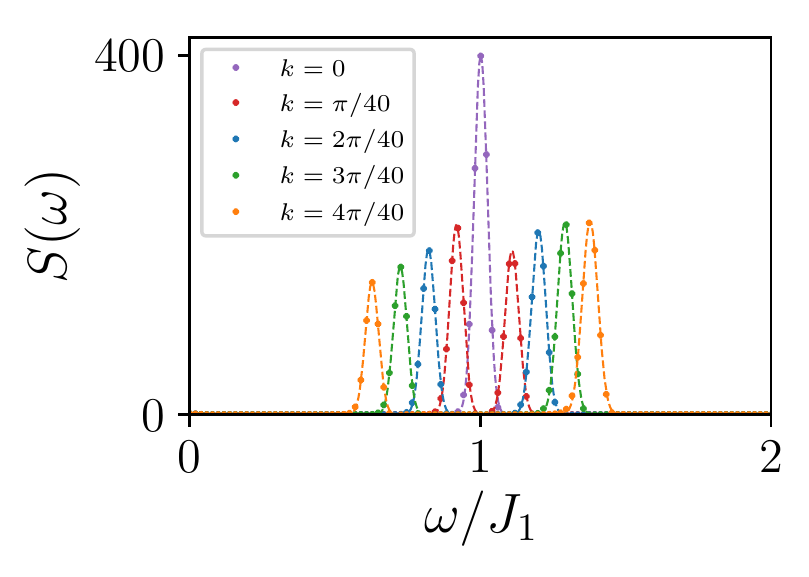} \put (0,67) {\footnotesize{(d)}} \end{overpic}
\caption{Dynamical correlations $S^{+-}(k,\omega)$ obtained numerically for Zeeman field of $B/J_1=1$. (a-b) $J_2=0$, (c-d) $J_2/J_1=0.24$. In (a,c) the red dashed line indicates a fit to the analytic expression in Eq.~\eqref{eq:chi_pm} valid in the vicinity of $k=0$. In (b,d) cuts of the dynamical correlations are shown for fixed values of $k$. }
    \label{fig:DynamicalCorrelationsSmallB}
\end{figure}

Now consider the interaction $g$.  Importantly, response at the energy of the order of the Zeeman energy $B$, see \eqref{eq:chi_pm_g0}, is determined 
by $g=g(B)\neq0$ in \eqref{eq:1} which is small and finite.
The diagonal $J_R^z J_L^z$
term in Eq.~\eqref{eq:1} leads for $M>0$ to a simple increase of the spin
splitting of the spinon bands by the energy $\Delta=gM/2$.  Consequently the full spin splitting is $B+\Delta$ and
na\"ively the poles in Eq.~\eqref{eq:chi_pm_g0} would be shifted
vertically to $B+\Delta \pm vk$.  This clearly violates the Larmor
theorem.  The contradiction is resolved by including the spin flip
part of the interaction $J_R^+J_L^- + \rm{h.c.}$, which results in the formation of a bound
state between the particle and hole (exciton) created by the spin
operator $S^+$.  The two effects together are captured by a Random
Phase Approximation summation of ladder diagrams for the
susceptibility, as described in \cite{SM}, leading to the result
\begin{eqnarray}\label{eq:chi_pm}
\chi^\pm(k,\omega) & = &  M \left( \frac{A_+(k)}{\omega-\omega_+(k)}+\frac{A_-(k)}{\omega-\omega_-(k)}\right) ,\\
A_{\pm}(k) & = & 1 \pm \frac{ \tilde{v}^2 k^2 - B\Delta }{  B\sqrt{\Delta^2 + \tilde{v}^2 k^2} }, \nonumber \\
\omega_{\pm}(k) & = & B + \Delta \pm \sqrt{\Delta^2+ \tilde{v}^2 k^2} . \nonumber
\end{eqnarray}
Here  $\tilde{v}=v\sqrt{1- g^2\chi_0^2/4} $.
This is plotted schematically in Fig.~\ref{fig:chi_pm_schematic}.  The
downward branch $\omega_-(k)$ has finite residue which approaches
$2 M$ for $k \to 0$ and $\omega_-(k) \to B$, satisfying the Larmor
theorem.  The spectral weight of the upward branch $\omega_+(k)$  vanishes
quadratically $A_+(k)\propto \tilde{v}^2 k^2$ for $k \to 0$, when
$\omega_+(k) \to B + 2\Delta$.  Both branches scale linearly with $\tilde{v} k$
for sufficiently large momenta $\tilde{v} k \gg \Delta$.  
Within our low-energy approximation the $k=0$ gap between the two branches is given by $2\Delta = \omega_{+}(0) - \omega_{-}(0) = g M$. 
Higher order in $g$ and $B$ contributions can modify it.

\begin{figure}
    \centering
    \begin{overpic}[width=0.49\columnwidth]{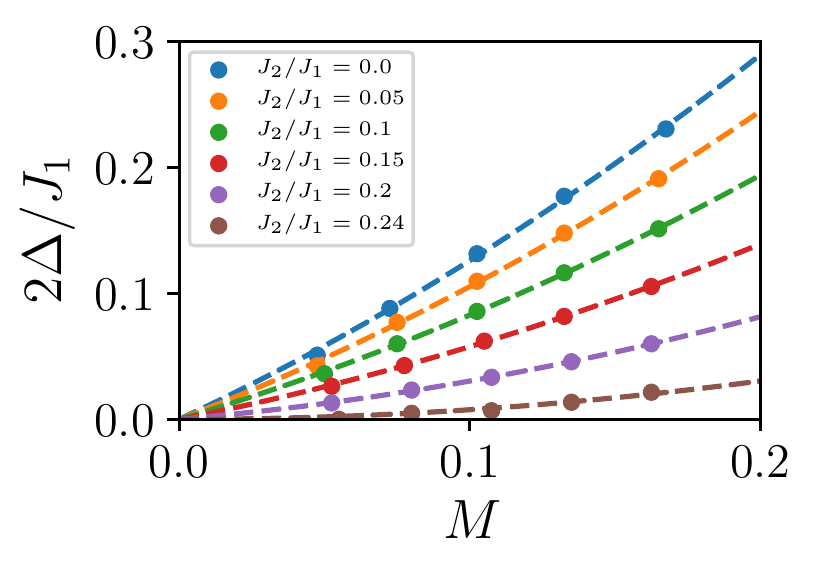} \put (0,65) {\footnotesize{(a)}} \end{overpic}
    \begin{overpic}[width=0.49\columnwidth]{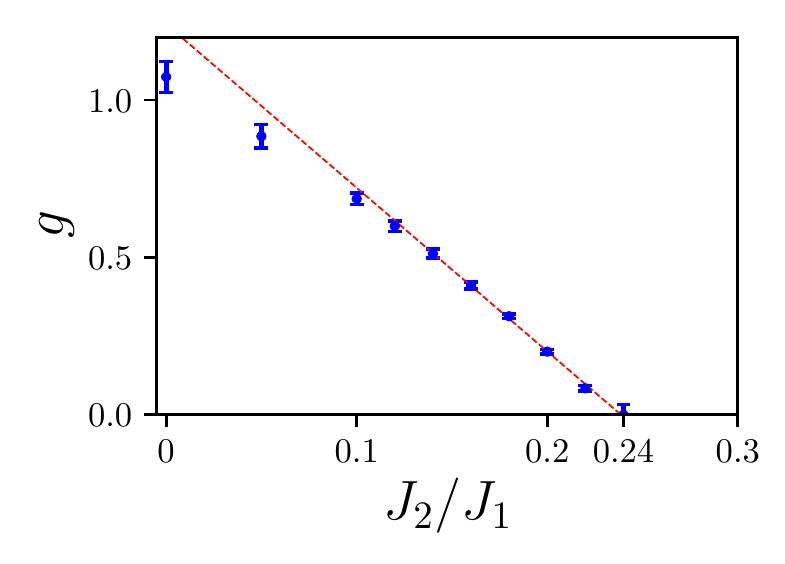} \put (0,65) {\footnotesize{(b)}} \end{overpic}
\caption{(a) The splitting at $k=0$ as function of the magnetization for different values of $J_2$ (b) The backscattering interaction $g$, extracted from $2\Delta$ vs $M$ in (a).}
    \label{fig:Splitting_vs_J2}
\end{figure}

We now compare our analytical analysis to numerical results, which are
consistent with earlier studies of the Heisenberg
chain\cite{Lefmann1996,Kohno2009,Nishimoto2007,Bouillot2011}.  These
works observed a finite gap, but did not address its origin
and systematics.  The dynamical correlations obtained numerically are
shown in Fig.~\ref{fig:DynamicalCorrelationsSmallB}.  Since the
spectral weight of the upper branch vanishes at $k=0$, to obtain the
gap $2\Delta$ we extract the dispersion at small momenta (see
Fig.~\ref{fig:DynamicalCorrelationsSmallB}(b,d)), fitting the two
branches to the form expected from Eq.~\eqref{eq:chi_pm}, and
extrapolating to $k=0$. The resulting gap versus magnetization, as
$J_2$ is varied, is shown in Fig.~\ref{fig:Splitting_vs_J2}(a).  We
account for higher order $M^2$ corrections to 
$\Delta$ by fitting the curves shown in
Fig.~\ref{fig:Splitting_vs_J2} to the form
$2\Delta=gM+\alpha M^2$, and extract $g(J_2)$ which is plotted in
Fig.~\ref{fig:Splitting_vs_J2}(b).  Additional data for ferromagnetic $J_2<0$, which
enhances $g$ beyond that of the nearest-neighbor limit, is given in \cite{SM}.
Extrapolating $g(J_2)$ to zero, we
find that $g$ vanishes at $J_2/J_1=0.239\pm 0.005$
in agreement with the critical value $J_{2,c}/J_1\approx 0.241$ \cite{Eggert1996} up to
numerical uncertainties.   Fixed momentum cuts of the $S^{+-}(k,\omega)$
(Fig.~\ref{fig:DynamicalCorrelationsSmallB}b,d) show that, as
predicted by Eq.~\eqref{eq:chi_pm}, the spectral weight of the upper
branch is  suppressed at small $k$ for the Heisenberg case (the
generic situation), while the two branches
have approximately equal weight in the free spinon limit ($J_2\approx J_{2,c}$).

\emph{High magnetization --}
We next consider the limit of a nearly polarized chain with low density of down spins, i.e. when the field is close to saturation value $B_{\rm sat} = (1+\eta)J_1$.
In this limit it is useful to consider the mapping of spins to spinless fermions defined by $S^-_i = \prod_{j<i} (-1)^{n_j} c^\dagger_i$ and $S^z=1/2-n_i$, where $c^\dagger_i$ denotes the fermionic creation operator on site $i$ and $n_i=c^\dagger_i c^\pd_i$. 
We focus on the case with $J_2=0$ first. The Hamiltonian \eqref{eq:H} maps to
\begin{equation}\label{eq:H_fermions}
H = \sum_i \frac{J_1}{2} \left( c^\dagger_i c^{\vphantom\dagger}_{i+1}
  + {\rm h.c.} \right) + \eta J_1  n_i n_{i+1} + (B-\eta J_1 ) n_i.
\end{equation}

\begin{figure}
    \centering
    \begin{overpic}[width=0.49\columnwidth]{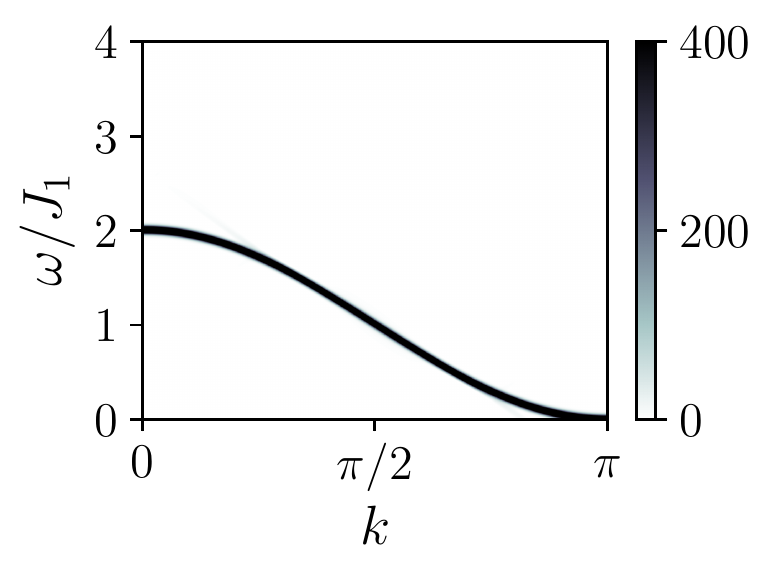} \put (0,68) {\footnotesize{(a)}} \end{overpic}
    \begin{overpic}[width=0.49\columnwidth]{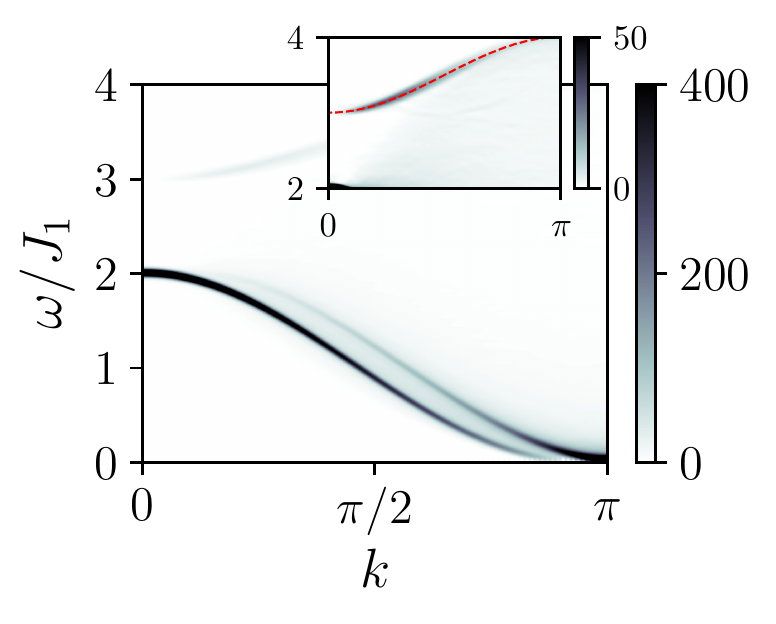} \put (0,68) {\footnotesize{(b)}} \end{overpic} \\
    \begin{overpic}[width=0.49\columnwidth]{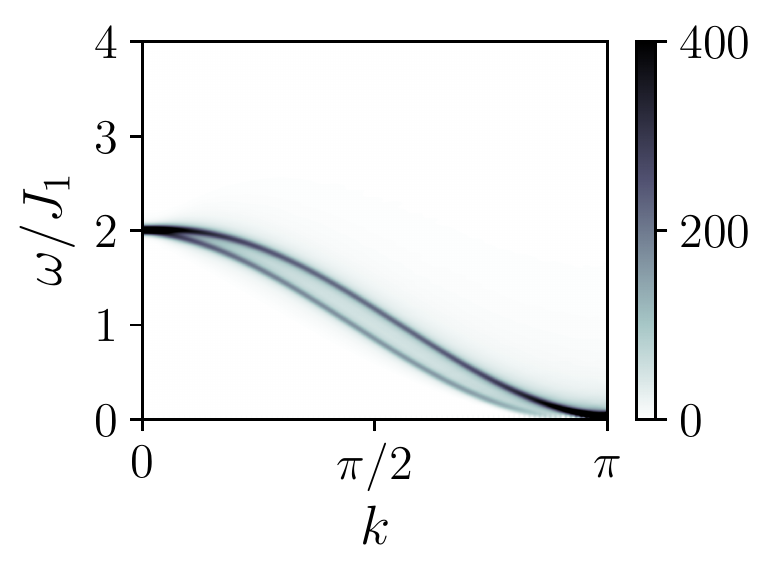} \put (0,68) {\footnotesize{(c)}} \end{overpic}
    \begin{overpic}[width=0.49\columnwidth]{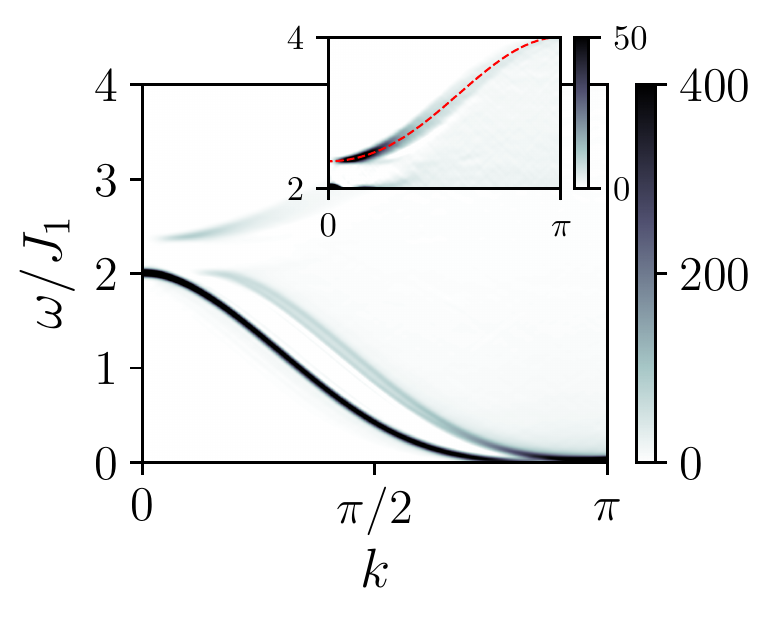} \put (0,68) {\footnotesize{(d)}} \end{overpic}
\caption{Dynamical correlations $S^{+-}(k,\omega)$ obtained using DMRG and time-evolution in the high magnetization regime. In (a-c) we set $J_2=0$. (a) Saturated chain, i.e. $M=1/2$, a pure cosine dispersion is observed. (b) $M=0.45$ for an isotropic chain, $\eta=1$, (c) $M=0.45$ in the non-interacting limit, $\eta=0$. (d) Isotropic chain $\eta=1$ for $J_2/J_1=0.24$. In (b,d) the inset shows the range $\omega>2J_1$ of the response function on a different color scale  that allows for better visibility of the high energy mode. The red dashed line corresponds to the two-magnon bound state dispersion $\epsilon_2(k+\pi)$.}
    \label{fig:DynamicalCorrelationsLargeB}
\end{figure}

At the saturation field $B=B_{\rm sat}$, the ground state is fully
polarized, $|0\rangle = |FM\rangle$, and the only contribution to the
dynamical susceptibility is the one-magnon state with momentum $k$,
$|1_k\rangle = \frac{1}{\sqrt{N}} \sum_m e^{ikm} S_m^- |FM\rangle$.
Consequently, the transverse correlations feature a sharp cosine mode
at $\omega=J_1 (1+ \cos k)$ as can be seen in
Fig.~\ref{fig:DynamicalCorrelationsLargeB}(a).   In
the isotropic case, i.e. for $\eta=1$, as the field is lowered and
the density of spin down particles increases, we observe a splitting
of the cosine mode as well as an appearance of a new mode at higher
energies $\omega>2J_1$ (see
Fig.~\ref{fig:DynamicalCorrelationsLargeB}(b)).  To understand this
response it is useful to compare to the limit of non-interacting fermions $\eta=0$.
The dynamical correlations
obtained in this limit are plotted in
Fig.~\ref{fig:DynamicalCorrelationsLargeB}(c).  It is seen that
the low energy response at $\omega<2J_1$ is not altered
significantly. Indeed, as shown in \cite{SM}, the splitting of the
lower mode can be understood in the non-interacting limit as
originating from single particle excitations above the Fermi
sea~\footnote{Note that in the non-interacting limit the many-body
  eigenstates are easily obtained as Slater-determinant states,
  however correlations remain non-trivial due to the presence of the
  string operator~\cite{Lieb1961} and can be calculated using the methods discussed in \cite{Bravyi2017}.}.  The mode at higher energies however is
completely gone for $\eta=0$, indicating that its presence comes
purely from interaction effects. In fact, for the Heisenberg chain, it
is known that this mode comes from Bethe ansatz string solutions,
which can be identified as two-magnon bound states close to the
saturation field~\cite{Muller1981,Kohno2009}. Here we 
show that it is a generic feature of the interacting magnons which exists beyond the integrable limit.

To examine two-particle bound state solutions we consider a state with two down spins
\begin{equation}
\left|2_K\right>=\sum_{m,n}\psi_{m,n}S^-_mS^-_n \left|FM\right>,\ \psi_{m,n}\propto e^{iK\left(\frac{m+n}{2}\right)} f(m-n).
\end{equation}
Due to translational invariance the two particle wavefunction $\psi_{m,n}$ can be written as above, with $K$ denoting the center of mass momentum of the pair of magnons. Looking for eigenstates of the Hamiltonian~\eqref{eq:H} of the form above, leads to an effective Schrodinger equation for $f(m-n)$. Requiring a bound solution for $f$
we can check that such a solution exists for a given $K$ and obtain its energy.
For $J_2=0$ the dispersion of the bound state can be easily obtained analytically and is given by $\epsilon_2(K, J_2=0)=2B-J_1\sin^2(K/2)$~\cite{Bethe1931}, while for finite $J_2$ we calculate the dispersion numerically \cite{SM}. 

To understand how the two-magnon bound states are revealed in the
transverse correlations, we consider for simplicity the limit of a single
down spin in the otherwise polarized ground state of length $N$.
This state is a
caricature of the many body ground state at a low density of spin flips
$n_{\rm sf} = 1/2-M=1/N \ll 1$ close to but below the saturation
field.  Note that since the  minimum of the single magnon dispersion
is at momentum $\pi$ for $J_1>0$ and $J_2/J_1<1/4$, the magnon present
in the ground state will occupy that momentum.   Hence we take
$|0\rangle = \left|1_{\pi}\right\rangle$ in Eq.~\eqref{eq:Spm}.  Now
there is a contribution when $\langle m| = \langle 2_K|$ in
Eq.~\eqref{eq:Spm}, which, by momentum conservation, gives a matrix
element 
$\left| \left\langle 2^\pd_K \right| S^{-}_k \left|1^\pd_\pi
  \right\rangle \right|^2 \propto \delta_{K,k+\pi}/N$ \cite{SM} (physically the
$1/N$ factor appears
because the  spin flip created by $S_k^-$ has only a small probability to occur
near the spin flip already present in the ground state).  This
implies the appearance of response at energy
$\omega=\epsilon_2(k+\pi)$ with an weight proportional to $1/N=1/2-M$ \cite{SM}. 
Plotting the expected dispersion due to the two-magnon bound state on
top of the dynamical correlations obtained numerically (dashed line in
Figs.~\ref{fig:DynamicalCorrelationsLargeB}(b,d)) we find an excellent
agreement between the two \cite{SM}.

Note that as opposed to the low-magnetization regime, where the
splitting between the modes at $k=0$ vanishes at $J_2 \to J_{2,c}$, in
the high-magnetization regime the splitting between the modes remains
finite, and from the aforementioned analysis is explicitly determined
by the two magnon bound state at $K=\pi$~\cite{Chubukov1991,Kecke2007} to be $2\Delta=J_1-3J_2+J_2^2/(J_1-J_2)$ \cite{SM}.
This highlights the fact that the nature and the origin of the high energy mode in the low and high magnetization regimes is very different. While in the low-magnetization regime the high energy mode describes a continuum of excitations and the low energy mode is a sharp collective excitation of spinons~\cite{BalentsStarykh2020}, in the high magnetization regime the situation is reversed: the low energy modes in the response from a continuum of psinon excitations \cite{karbach2002} while the high energy mode is a sharp two-magnon bound state.

\emph{Discussion--} 
Our study is complementary to a prior body of work on spectral
functions of one dimensional systems beyond conventional Luttinger
liquid theory\cite{RevModPhys.84.1253,PhysRevLett.100.027206}, which
discussed Heisenberg and related chains but focused on zero magnetic
field.  Other studies in small non-zero magnetic fields were motivated in
part by electron spin resonance.  The pioneering work of Oshikawa and
Affleck noted the irrelevance of backscattering at zero field, and
argued that the Larmor theorem shows that it has negligible effect in
small fields\cite{OshikawaAffleck2002}.  A later study by Karimi and
Affleck\cite{PhysRevB.84.174420} included non-linear terms in the
fermion dispersion, as well as the effect of the longitudinal part of the
backscattering interaction, but not the transverse interactions; hence
this misses the formation of the bound state at $k\rightarrow 0$.  

We are optimistic that these results might be observed in experiment.
Indeed there are a number of recent studies that observed spectral
features interpreted as Bethe string states via high-resolution
terahertz spectroscopy~\cite{Wang2018} and inelastic neutron
scattering \cite{Bera2020}. 
Earlier neutron scattering studies in
non-zero magnetic field~\cite{Heilmann1978,Stone2003} also contain
hints of the interaction signatures discussed here.  In an ultra-cold
atomic realization of a Heisenberg chain, bound states have been
observed by quite different real time protocols~\cite{Ganahl2012,Fukuhara2013}.
The implications of our
results for the spectral features at partial polarization to such real
time experiments is an interesting direction for future studies. 

\acknowledgements \emph{Acknowledgements-} We would like to thank
  R. Coldea and L. Motrunich for inspiring remarks and questions,
  J. S. Caux for pointing out Ref.~\cite{Bouillot2011}, and
  M. Kohno for discussions of Ref.~\cite{Kohno2009}. This research
is funded in part by the Gordon and Betty Moore Foundation through
Grant GBMF8690 to UCSB to support the work of A.K. Use was made of the
computational facilities administered by the Center for Scientific
Computing at the CNSI and MRL (an NSF MRSEC; DMR-1720256) and
purchased through NSF CNS-1725797.  This work was supported by the NSF CMMT
program under grant No.  DMR-1818533 (L.B.) and DMR-1928919 (O.A.S.).
We benefitted from the facilities of the KITP, NSF grant PHY-1748958.

\bibliography{refs}

\clearpage

\onecolumngrid
\setcounter{equation}{0}
\setcounter{figure}{0}
\setcounter{secnumdepth}{3}
\renewcommand{\theequation}{S\arabic{equation}}
\renewcommand{\thefigure}{S\arabic{figure}}

\begin{center}
{\Large\bfseries Supplementary Material}
\end{center}

\section{Additional numerical results}

Numerical calculations are performed on a finite chain of length $N=400$ sites with open boundary conditions. Denoting the ground state by $| 0 \rangle$,
the real-time correlation function can be written as
\begin{equation}
 \left \langle 0 \right| S_j^{+}(t) S_0^{-}(0) \left| 0\right\rangle= e^{i E_{0} t} \left\langle 0 \right| S_j^{+} e^{-i H t} S_0^{-} \left| 0 \right\rangle,
\end{equation}
where $E_0$ is the ground state energy, and the site index $0$ refers to the site in the middle of the chain. The real-time correlation function can be obtained by time evolving the state $| 1 \rangle \equiv S^{-}_0|0\rangle$ and calculating the respective overlap. 
We perform the time evolution using TEBD, employing a 4th order Suzuki-Trotter decomposition with a time step of $dt=0.005 J_1^{-1}$. In presence of a finite $J_2$ pairs of spin-$1/2$ sites are grouped into supersites, and a standard TEBD algorithm is used for the supersites. Setting the maximal MPS bond dimension to $M=200$, we find that the maximal truncated weight $\epsilon_{\rm tr}$ in our simulations remains below $\sim10^{-7}$ in the low-magnetization regime and below $\sim10^{-8}$ in the high-magnetization regime with $M \geq 0.4$. Performing the time evolution up to times $t_{\rm max} = 80 J_1^{-1}$, we verify that the correlations grow to a finite value only in the bulk of the system and the excitations do not reach the boundary. 
Due to the even number of sites in the chain we perform symmetrization of the correlations with respect to the site $j=0$.
To further improve the frequency resolution, we use linear prediction~\cite{White2008}. To this end, we first perform the spatial Fourier transform of the spin-spin correlation function, and use linear prediction to extrapolate the correlation functions in momentum space up to times $2t_{\rm max}$. We then apply a Gaussian windowing function $\exp[-t^2/(0.75 t_{\rm max})^2]$, and finally perform the time Fourier transform to obtain $S^{+-}(k,\omega)$. 

\begin{figure}[b]
\includegraphics[width=0.95\textwidth]{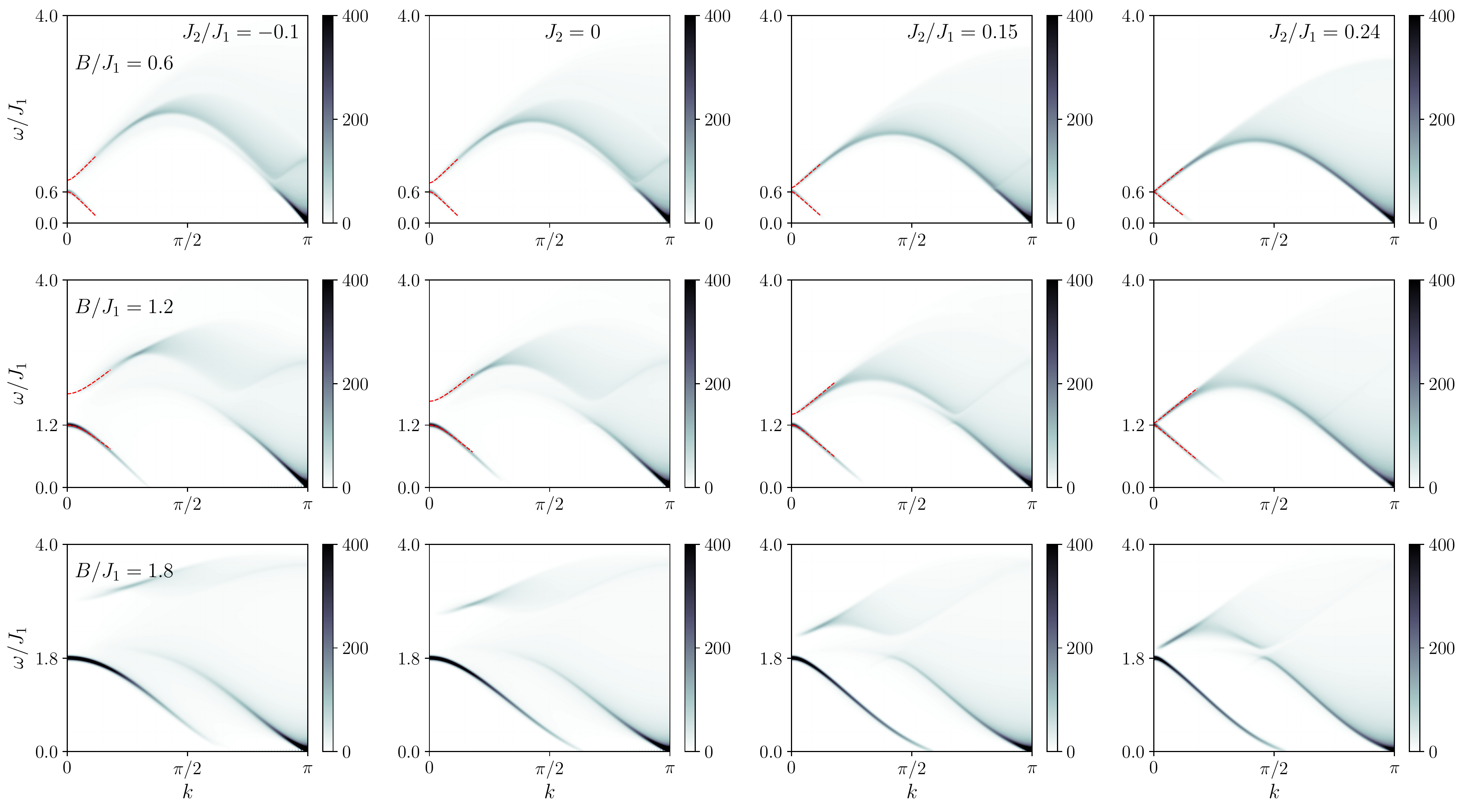}
\caption{Dynamical correlations $S^{+-}(k,\omega)$ obtained numerically as the coupling $J_2$ and Zeeman field $B$ are varied for an isotropic chain with $\eta=1$. The red dashed line indicates a fit to the analytic expression in Eq.~\eqref{eq:chi_pm} of the main text valid in the vicinity of $k=0$ in the low magnetization regime.}
\label{fig_sm:SpmLowMag}
\end{figure}

\begin{figure}[t]
\includegraphics[width=0.95\textwidth]{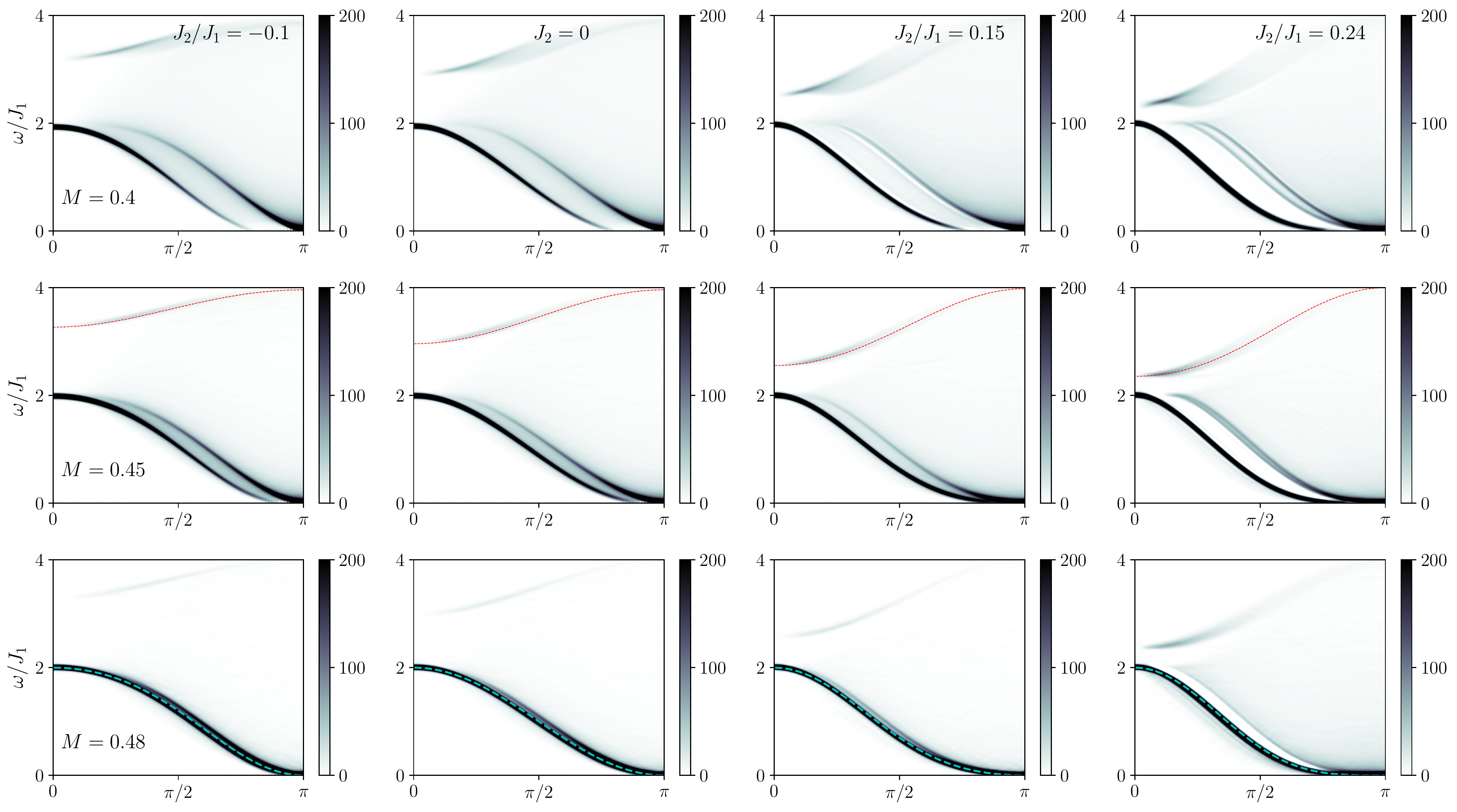}
\caption{Dynamical correlations $S^{+-}(k,\omega)$ obtained numerically as the coupling $J_2$ and the magnetization $M$ are varied for an isotropic chain with $\eta=1$. The red dashed line plotted on top of the plots obtained for $M=0.45$ corresponds to the two-magnon bound state dispersion $\epsilon_2(k+\pi)$ (see main text). The cyan dashed line plotted on top of the plots obtained for $M=0.48$ corresponds to the single magnon dispersion $\epsilon_1(k)=J_1[\cos(k) -\eta]+J_2[\cos(2 k) -\eta]+B$.} 
\label{fig_sm:SpmHighMag}
\end{figure}

\begin{figure}[b]
\includegraphics[width=0.4\textwidth]{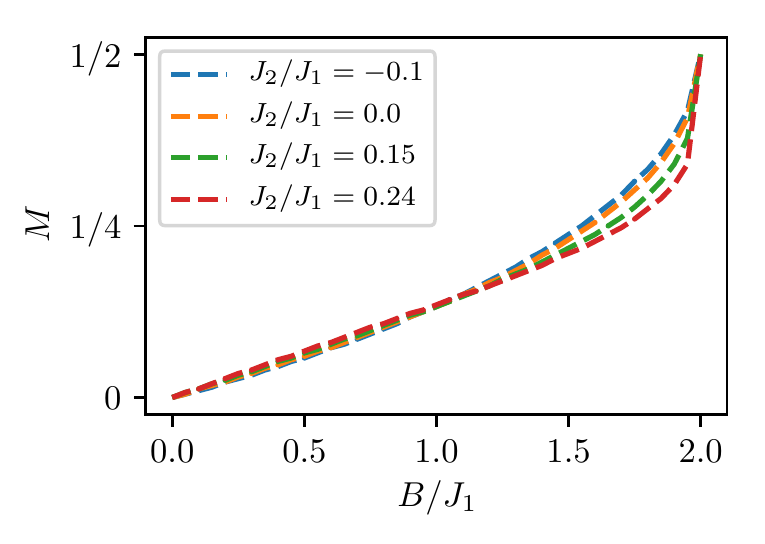}
\caption{Magnetization of the chain as function of the Zeeman field for an isotropic chain with $\eta=1$ for different values of $J_2/J_1$. }
\label{fig_sm:magnetization}
\end{figure}

In the main text we show the transverse dynamical correlations obtained for $J_2=0$ and $J_2/J_1=0.24$, i.e. close to the critical value, for a single value of the longitudinal Zeeman field $B$ in the low-magnetization regime, and for a single magnetization $M$ in the high-magnetization regime. Here, in Figs.~\ref{fig_sm:SpmLowMag} and \ref{fig_sm:SpmHighMag} we present additional results as the values of $J_2$ and $B$ or $M$ are varied. 
In addition, in Fig.~\ref{fig_sm:magnetization} we show the magnetization of the chain as function of the field for different values of $J_2/J_1$.

\clearpage
\section{Dynamic susceptibility in the low magnetization regime}

Heisenberg chain is described by spin-1/2 Dirac fermions $\psi_{R,s}, \psi_{L,s}$ subject to backscattering interaction $V = - g \int dx \vec{J}_R \cdot \vec{J}_L$. Here the right spin current $\vec{J}_R = \frac{1}{2} \psi_R^\dagger \vec{\sigma} \psi_R$ and similarly for the left current. The notation is that $\psi_R$ without spin subindex $s$ denotes two-component spinor $\psi_R = (\psi_{R \uparrow}, \psi_{R\downarrow})^{\rm T}$. The full Hamiltonian is $H = H_0 + H_B + V$ where the free part $H_0$ describes fermion kinetic energy and 
$H_B$ describes Zeeman interaction with the magnetic field $B$
\be
H_0 + H_B = v \int dx \Big( \psi_R^\dagger (-i \partial_x) \psi_R + \psi_L^\dagger (i \partial_x) \psi_L \Big) - B (J^z_R + J^z_L)
\label{eq:3}
\ee
Note that magnetization operator $\hat{M} = J^z_R + J^z_L$ is just the sum of the right and left spin currents. Its expectation value is magnetization itself,
$M = \langle J^z_R + J^z_L\rangle$.

The backscattering interaction is broken into transverse and longitudinal parts, $V = V_1 + V_2$, where
\bea
V_1 &=& -\frac{g}{2} \int dx \Big( J^+_R J_L^- + J_R^- J_L^+ \Big) =  -\frac{g}{2} \int dx ~ \Big( \psi_{R\uparrow}^\dagger \psi_{R\downarrow} \psi_{L\downarrow}^\dagger \psi_{L\uparrow} + \psi_{R\downarrow}^\dagger \psi_{R\uparrow} \psi_{L\uparrow}^\dagger \psi_{L\downarrow}  \Big)
\label{eq:5}
\eea
and
\bea
V_2 &=& - g \int dx J^z_R J^z_L = -\frac{g}{4} \sum_{s,s'} s s' \int dx ~\psi_{R s}^\dagger \psi_{R s} \psi_{L{s'}}^\dagger \psi_{L{s'}} .
\label{eq:6}
\eea
As is well-known, backscattering interaction constant $g$ flows under RG transformation as 
\be
g(\ell) = \frac{g(0)}{1+ g(0) \ell}
\label{eq:7}
\ee
where $g(0)$ is its initial value at the lattice scale and $\ell=\ln(J/E)$ is the logarithmic RG scale, where $J\approx J_1$ is the high-energy (lattice scale cut-off) and $E$ is the running energy. Without the magnetic field, $g(\ell) \to 1/\ell$ vanishes as energy goes to zero, $E\to 0$. In the presence of the field, however, this marginally irrelevant flow is terminated at
$E = B$, at which the transverse backscattering $V_1$ effectively averages to $0$ and the RG flow stops \cite{affleckoshikawa1999}. For energy below this critical value, the coefficient of $V_2$ term is 
given by the finite constant $g_z \approx g(\ell_b)$, with $\ell_b = \ln(J/B)$, while that of the transverse term $V_1$ turns to zero, $g_\perp =0$. Importantly, at $E = B$ both coupling constants are finite and given by $g = g(\ell_b)$ \cite{affleckoshikawa1999}. This is the essence the Kosterlitz-Thouless RG flow for the spin-1/2 chain in magnetic field, see Fig.6 in \cite{affleckoshikawa1999} and discussion therein. It is this finite and constant $g = g(\ell_b)$ that shows up in our calculation, simply because we are studying dynamic response of the spin chain at the Zeeman energy $B$,
and not at $E \approx 0$ as is often done when one considers the response at the lowest possible energy.

The longitudinal interaction produces ``exchange field" $g M/2$ as can be easily seen with the help of a standard mean-field approximation 
\bea
V_2 &=& - g \int dx J^z_R J^z_L \approx - g \int dx \big(\langle  J^z_R\rangle  J^z_L + \langle  J^z_L\rangle  J^z_R\big) = -\frac{1}{2} g M \int dx \big(  J^z_R +   J^z_L\big),
\label{eq:8}
\eea
where we used the fact that magnetization $M$ is symmetrically split between the right- and left-movers, $\langle J^z_R\rangle = \langle J^z_L\rangle = M/2$.
Therefore the total field experienced by spinons is $B_{\rm tot} = B + g M/2$. Magnetization $M$ is the response to $B_{\rm tot}$, so that
\be
M = \chi_0 B_{\rm tot} = \chi_0 (B + \frac{1}{2} g M),
\label{eq:9}
\ee
where $\chi_0 = 1/(2\pi v)$ is the susceptibility of non-interacting gas of spinons. The static susceptibility $\chi$ of interacting spinons then follows as
\be
M = \frac{\chi_0}{1-\frac{1}{2} g \chi_0} B = \chi B .
\label{eq:10}
\ee
For positive $g>0$ $\chi$ is enhanced by interactions relative to $\chi_0$ and is generic feature of interacting fermions, i.e. of the Fermi-liquid.

The same result for the ``exchange" addition to $B$ can be obtained diagrammatically, by considering tadpole self-energy diagram produced by $V_2$ to the fermion Green's function of right/left-movers with spin projection $s$. Calculating the transverse spin susceptibility as a convolution of the spin-up and spin-down Green's functions one finds that the external field is modified by the exchange correction, $B \to B + g M/2$. See \cite{BalentsStarykh2020} for similar considerations in two spatial dimensions.

We now focus on the uniform part of the transverse spin susceptibility $\chi^\pm(k,\omega)$, with $k \approx 0$. This is given by the Fourier transform of the retarded Green's function of the spin currents, $\chi^\pm_{\rm ret}(x,t) = - i \Theta(t) \langle [J_R^+(x,t) + J_L^+(x,t), J_R^-(0) + J_L^-(0)]\rangle$. To find this object, we turn to the Matsubara Green's function 
\be
G_{\mu\nu}(x,\tau) = - \langle \hat{T}_\tau J_\mu^+(x,\tau) J_\nu^-(0,0)\rangle
\label{eq:11}
\ee
where indices $\mu, \nu \in (R,L)$ denote right and left spin
currents. Angular brackets denote average with respect to  the
action for interacting spinons ${\cal S} = {\cal S}_0 + {\cal S}_{\rm int}$, where we take ${\cal S}_0$ to represent action of the free right- and left-moving fermions subject to the {\em total} magnetic field $B_{\rm tot} = B + g M/2$. The interaction part 
${\cal S}_{\rm int} = -\int \!d\tau\, V_1$ describes backscattering
interaction between transverse components of spin currents of the
opposite chirality. It is easy to see that in the non-interacting
theory (one without $V_1$) there are no correlations between the right
and left spin currents, $G^0_{R L}(x,\tau) = G^0_{L R}(x,\tau)
=0$. However finite ${\cal H}_{\rm int}$ produces finite $G_{R L}$ and
$G_{L R}$.

To calculate it, we employ an RPA approximation.  It is most compact
to express this in a path integral formulation.  As we are interested
only in correlations of the currents, we can formulate a {\em bosonic}
path integral by introducing composite variables via a functional
delta function.  The weight $W$, where
\begin{equation}
  \label{eq:2}
  W[J^\mu_R,J^\mu_L] = \int D[\psi_{R/L}]
  e^{S_0+S_{\rm int}}\delta[J_R^\mu-\bar\psi_R
  \frac{\sigma^\mu}{2}\psi_R] \delta[J_L^\mu-\bar\psi_L
  \frac{\sigma^\mu}{2}\psi_L] ,
\end{equation}
by construction is a functional of the bosonic fields
$J_R^\mu,J_L^\mu$ which reproduces the correlation functions of the corresponding
currents defined in the fermionic theory, if used as a statistical
weight in the path integral.  It is convenient to define a generating
function
\begin{equation}
  \label{eq:4}
  Z[Q^+,Q^-] = \int D[J_{R/L}] W[J_R,J_L] e^{  \int\! d\tau dx\,
    \left[ Q^+(J_R^-+J_L^-) + Q^-(J_R^++J_L^+)\right]}.
\end{equation}
Here $Q^\pm$ are infinitesimal c-number boson fields which allow us to
obtain the desired correlations, specifically
\begin{equation}
  \label{eq:26}
  G =- \left\langle \hat{T}_\tau (J_R^+ +
    J_L^+)_{x,\tau}(J_R^-+J_L^-)_{0,0}\right\rangle = - \left.\frac{\delta^2
    \ln Z}{\delta Q^+(0,0) \delta Q^-(x,\tau)}\right|_{Q^\pm=0}.
\end{equation}
Now because of the functional delta function in the definition in
Eq.~\eqref{eq:2}, one can express the interaction term directly in
terms of the bosonic fields, so that
\begin{equation}
  \label{eq:27}
  W[J^\mu_R,J^\mu_L] = e^{\frac{g}{2} \int \!  dx d\tau\, (J_R^+ J_L^-
    + J_R^- J_L^+)}\int D[\psi_{R/L}]
  e^{S_0}\delta[J_R^\mu-\bar\psi_R
  \frac{\sigma^\mu}{2}\psi_R] \delta[J_L^\mu-\bar\psi_L
  \frac{\sigma^\mu}{2}\psi_L] .
\end{equation}
The Grassman integral remaining here is the statistical weight defined
for the non-interacting fermion WZW theory.  Because the currents as
operators are quadratic in the fermions, this is not a Gaussian
functional, but the RPA approximation consists  in regarding it as
one.  That is, we take the Grassman integral to be approximated as the
exponential of a quadratic form in the currents.  The coefficients are
determined by requiring that, for the free theory, this quadratic form
reproduces the known current-current correlators of the WZW theory.
Henceforth we drop terms involving $J_{R/L}^z$, because these are not
needed in the calculation.  One then has
\begin{equation}
  \label{eq:28}
  W[J_R,J_L] = e^{S_{\rm eff}[J_R^\mu,J_L^\mu]},
\end{equation}
where
\begin{equation}
  \label{eq:29}
  S_{\rm eff} = \int\! \frac{dk}{2\pi} \frac{d\omega_n}{2\pi}\, \left[ \left(G_{RR}^0\right)^{-1}
    J_R^+ J_R^- + \left(G_{LL}^0\right)^{-1}
    J_L^+ J_L^- + \frac{g}{2} \left( J_R^+ J_L^- + J_R^+ J_L^-\right)\right].
\end{equation}
From this point on it is convenient to work in Fourier space because
the Green's functions are diagonal.
Now we seek from Eq.~\eqref{eq:4}
\begin{equation}
  \label{eq:35}
  Z[Q^+,Q^-] = \int D[J_{R/L}]  e^{\tilde{S}_{\rm eff}(J,Q)},
\end{equation}
where
\begin{equation}
  \label{eq:36}
  \tilde{S}_{\rm eff} = \int\! \frac{dk}{2\pi}
  \frac{d\omega_n}{2\pi}\, \left[ J^+\cdot {\sf M}\cdot J^- + Q^+ J^-
  \cdot \begin{pmatrix} 1 \\ 1\end{pmatrix} + Q^- J^+\cdot \begin{pmatrix} 1 &
    1\end{pmatrix} \right],
\end{equation}
where we introduced $J^+ = \begin{pmatrix} J_R^+ & J_L^+\end{pmatrix}$
and $J^- = \begin{pmatrix} J_R^- \\ J_L^-\end{pmatrix}$, and the
matrix
\begin{equation}
  \label{eq:37}
  {\sf M} = \begin{pmatrix} \left(G_{RR}^0\right)^{-1} & \frac{g}{2}
    \\ \frac{g}{2} & \left(G_{LL}^0\right)^{-1}\end{pmatrix}.
\end{equation}
The $Q$ dependence is extracted as usual by completing the square,
\begin{equation}
  \label{eq:38}
J^+  \rightarrow J^+ - Q^+ \begin{pmatrix} 1 &
    1\end{pmatrix} \cdot {\sf M}^{-1}, \qquad J^-  \rightarrow J^- -
  Q^-\, {\sf M}^{-1}\cdot \begin{pmatrix} 1 \\
    1\end{pmatrix}.
\end{equation}
This gives
\begin{equation}
  \label{eq:39}
  Z[Q^+,Q^-] = Z_0\exp\left[- \int\! \frac{dk}{2\pi} \frac{d\omega_n}{2\pi}\, Q^+ Q^- \begin{pmatrix} 1 &
    1\end{pmatrix} \cdot {\sf M}^{-1}\cdot \begin{pmatrix} 1 \\
    1\end{pmatrix}\right],
\end{equation}
where $Z_0$ is independent of $Q^\pm$.  Using Eq.~\eqref{eq:26}, and
computing the matrix element of ${\sf M}^{-1}$ in Eq.~\eqref{eq:39},
we obtain finally
\begin{equation}
G(k,\omega_n) =  \frac{G^0_{R R}(k,\omega_n) + G^0_{LL}(k,\omega_n) - g G^0_{RR}(k,\omega_n)G^0_{LL}(k,\omega_n)}{1-\frac{g^2}{4} G^0_{RR}(k,\omega_n)G^0_{LL}(k,\omega_n)} .
\label{eq:20}
\end{equation}

This result is easy to understand diagrammatically. Indeed, it is written in terms of the interaction vertex $g$ and bare Green's functions $G^0_{RR/LL}$, 
which we represent graphically as
\begin{equation}
  \label{oo2}
G^0_{RR}(k,i\omega_n)  =
                               \begin{tikzpicture}%[baseline={(current bounding box.center)}]
    % coordinates
     \tikzset{
      fermion/.style={thin,postaction={decorate},
        decoration={markings,mark=at position .7 with {\arrow[]{>}}}},
         fermion2/.style={thin,color=red,postaction={decorate},
        decoration={markings,mark=at position .7 with {\arrow[]{>}}}},
        fermion3/.style={thin,color=blue,postaction={decorate},
        decoration={markings,mark=at position .7 with {\arrow[]{>}}}},}
   \coordinate (x1) at (0,1.2);
   \coordinate (x2) at (1,1.2);
        \draw [fermion2] (x1) to (x2);
 \end{tikzpicture} , \,
 G^0_{LL}(k,i\omega_n)  =
                               \begin{tikzpicture}%[baseline={(current bounding box.center)}]
    % coordinates
     \tikzset{
      fermion/.style={thin,postaction={decorate},
        decoration={markings,mark=at position .7 with {\arrow[]{>}}}},
         fermion2/.style={thin,color=red,postaction={decorate},
        decoration={markings,mark=at position .7 with {\arrow[]{>}}}},
        fermion3/.style={thin,color=blue,postaction={decorate},
        decoration={markings,mark=at position .7 with {\arrow[]{>}}}},}
   \coordinate (x1) at (0,1.2);
   \coordinate (x2) at (1,1.2);
        \draw [fermion3] (x1) to (x2);
 \end{tikzpicture} , \,
 \frac{g}{2} =   \begin{tikzpicture}%[baseline={(current bounding box.center)}]
  \draw (0,1.2) circle (2pt);
  \end{tikzpicture}
\end{equation}
The total response \eqref{eq:26} is given by the sum of the renormalized Green's functions $G_{RR}, G_{LL}, G_{RL}$ and $G_{LR}$. Each of these
is given by a simple RPA-type diagram series
\begin{eqnarray}
  \label{oo3}
&&  G_{RL}(k,i\omega_n)  =
                               \begin{tikzpicture}%[baseline={(current bounding box.center)}]
    % coordinates
     \tikzset{
      fermion/.style={thin,postaction={decorate},
        decoration={markings,mark=at position .7 with {\arrow[]{>}}}},
         fermion2/.style={thin,color=red,postaction={decorate},
        decoration={markings,mark=at position .7 with {\arrow[]{>}}}},
        fermion3/.style={thin,color=blue,postaction={decorate},
        decoration={markings,mark=at position .7 with {\arrow[]{>}}}},}
   \coordinate (x1) at (0,1.2);
   \coordinate (x3) at (2,1.2);
       \draw (1,1.2) circle (2pt) coordinate (x2);
       \draw [fermion2] (x1) to (x2);
       \draw [fermion3] (x2) to (x3);
 \end{tikzpicture}
+\cdots , \,
 G_{LR}(k,i\omega_n)  =
                               \begin{tikzpicture}%[baseline={(current bounding box.center)}]
    % coordinates
      \tikzset{
      fermion/.style={thin,postaction={decorate},
        decoration={markings,mark=at position .7 with {\arrow[]{>}}}},
         fermion2/.style={thin,color=red,postaction={decorate},
        decoration={markings,mark=at position .7 with {\arrow[]{>}}}},
        fermion3/.style={thin,color=blue,postaction={decorate},
        decoration={markings,mark=at position .7 with {\arrow[]{>}}}},}
   \coordinate (x1) at (0,1.2);
   \coordinate (x3) at (2,1.2);
     \draw (1,1.2) circle (2pt) coordinate (x2);
      \draw [fermion3] (x1) to (x2);
      \draw [fermion2] (x2) to (x3);
   \end{tikzpicture}
+\cdots , \\
&& G_{RR}(k,i\omega_n)  =
\begin{tikzpicture}%[baseline={(current bounding box.center)}]
    % coordinates
     \tikzset{
      fermion/.style={thin,postaction={decorate},
        decoration={markings,mark=at position .7 with {\arrow[]{>}}}},
         fermion2/.style={thin,color=red,postaction={decorate},
        decoration={markings,mark=at position .7 with {\arrow[]{>}}}},
        fermion3/.style={thin,color=blue,postaction={decorate},
        decoration={markings,mark=at position .7 with {\arrow[]{>}}}},}
   \coordinate (x1) at (0,1.2);
   \coordinate (x2) at (1,1.2);
   \coordinate (x3) at (2,1.2);
        \draw [fermion2] (x1) to (x2);
 \end{tikzpicture}
+
                               \begin{tikzpicture}%[baseline={(current bounding box.center)}]
    % coordinates
     \tikzset{
      fermion/.style={thin,postaction={decorate},
        decoration={markings,mark=at position .7 with {\arrow[]{>}}}},
         fermion2/.style={thin,color=red,postaction={decorate},
        decoration={markings,mark=at position .7 with {\arrow[]{>}}}},
        fermion3/.style={thin,color=blue,postaction={decorate},
        decoration={markings,mark=at position .7 with {\arrow[]{>}}}},}
   \coordinate (x1) at (0,1.2);
   \coordinate (x3) at (2,1.2);
   \coordinate (x4) at (3,1.2);
     \draw (1,1.2) circle (2pt) coordinate (x2);      
     \draw [fermion2] (x1) to (x2);
      \draw [fermion3] (x2) to (x3);
          \draw (2,1.2) circle (2pt) coordinate (x3); 
          \draw [fermion2] (x3) to (x4);
\end{tikzpicture}
+\cdots , \,
 G_{LL}(k,i\omega_n)  =
\begin{tikzpicture}%[baseline={(current bounding box.center)}]
    % coordinates
     \tikzset{
      fermion/.style={thin,postaction={decorate},
        decoration={markings,mark=at position .7 with {\arrow[]{>}}}},
         fermion2/.style={thin,color=red,postaction={decorate},
        decoration={markings,mark=at position .7 with {\arrow[]{>}}}},
        fermion3/.style={thin,color=blue,postaction={decorate},
        decoration={markings,mark=at position .7 with {\arrow[]{>}}}},}
   \coordinate (x1) at (0,1.2);
   \coordinate (x2) at (1,1.2);
   \coordinate (x3) at (2,1.2);
        \draw [fermion3] (x1) to (x2);
 \end{tikzpicture}
+
                               \begin{tikzpicture}%[baseline={(current bounding box.center)}]
    % coordinates
     \tikzset{
      fermion/.style={thin,postaction={decorate},
        decoration={markings,mark=at position .7 with {\arrow[]{>}}}},
         fermion2/.style={thin,color=red,postaction={decorate},
        decoration={markings,mark=at position .7 with {\arrow[]{>}}}},
        fermion3/.style={thin,color=blue,postaction={decorate},
        decoration={markings,mark=at position .7 with {\arrow[]{>}}}},}
   \coordinate (x1) at (0,1.2);
   \coordinate (x3) at (2,1.2);
   \coordinate (x4) at (3,1.2);
     \draw (1,1.2) circle (2pt) coordinate (x2);      
     \draw [fermion3] (x1) to (x2);
      \draw [fermion2] (x2) to (x3);
          \draw (2,1.2) circle (2pt) coordinate (x3); 
          \draw [fermion3] (x3) to (x4);
\end{tikzpicture}
+\cdots 
\end{eqnarray}
It is also easy to see that the obtained equations are of Dyson type. For example, denoting the renormalized Green's function $G_{LL}$ by a thick blue line we can re-write
the last equation above as
\begin{equation}
G_{LL}(k,i\omega_n) = \begin{tikzpicture}%[baseline={(current bounding box.center)}]
    % coordinates
     \tikzset{
      fermion/.style={thin,postaction={decorate},
        decoration={markings,mark=at position .7 with {\arrow[]{>}}}},
         fermion2/.style={thin,color=red,postaction={decorate},
        decoration={markings,mark=at position .7 with {\arrow[]{>}}}},
        fermion4/.style={thick,color=blue,postaction={decorate},
        decoration={markings,mark=at position .7 with {\arrow[]{>}}}},}
   \coordinate (x1) at (0,1.2);
   \coordinate (x2) at (1,1.2);
   \coordinate (x3) at (2,1.2);
        \draw [fermion4] (x1) to (x2);
 \end{tikzpicture}
 = 
\begin{tikzpicture}%[baseline={(current bounding box.center)}]
    % coordinates
     \tikzset{
      fermion/.style={thin,postaction={decorate},
        decoration={markings,mark=at position .7 with {\arrow[]{>}}}},
         fermion2/.style={thin,color=red,postaction={decorate},
        decoration={markings,mark=at position .7 with {\arrow[]{>}}}},
        fermion3/.style={thin,color=blue,postaction={decorate},
        decoration={markings,mark=at position .7 with {\arrow[]{>}}}},}
   \coordinate (x1) at (0,1.2);
   \coordinate (x2) at (1,1.2);
   \coordinate (x3) at (2,1.2);
        \draw [fermion3] (x1) to (x2);
 \end{tikzpicture}
+
\begin{tikzpicture}%[baseline={(current bounding box.center)}]
    % coordinates
     \tikzset{
      fermion/.style={thin,postaction={decorate},
        decoration={markings,mark=at position .7 with {\arrow[]{>}}}},
         fermion2/.style={thin,color=red,postaction={decorate},
        decoration={markings,mark=at position .7 with {\arrow[]{>}}}},
        fermion3/.style={thin,color=blue,postaction={decorate},
        decoration={markings,mark=at position .7 with {\arrow[]{>}}}},
         fermion4/.style={thick,color=blue,postaction={decorate}, decoration={markings,mark=at position .7 with {\arrow[]{>}}}},}
   \coordinate (x1) at (0,1.2);
   \coordinate (x3) at (2,1.2);
   \coordinate (x4) at (3,1.2);
     \draw (1,1.2) circle (2pt) coordinate (x2);      
     \draw [fermion3] (x1) to (x2);
      \draw [fermion2] (x2) to (x3);
          \draw (2,1.2) circle (2pt) coordinate (x3); 
          \draw [fermion4] (x3) to (x4);
\end{tikzpicture}
\end{equation}
This, of course, is easy to solve
\be
G_{LL}(k,i\omega_n) = \frac{G^0_{LL}(k,\omega_n)}{1-\frac{g^2}{4} G^0_{RR}(k,\omega_n)G^0_{LL}(k,\omega_n)} .
\ee
Other renormalized Green's functions can be found similarly.

Next, we analytically continue \eqref{eq:20} to real frequencies, $\omega_n \to \omega + i 0$, and turn it into the expression for retarded transverse susceptibility:
$G^0_{RR/LL}(k,\omega + i 0) = \chi^\pm_{0,RR/LL}(k, \omega)$ and $G(k,\omega + i 0) = \chi^\pm(k,\omega)$.
The non-interacting susceptibility is given by 
\bea
\chi^\pm_{0,R R}(k, \omega) &=& \frac{1}{N} \sum_{k_1} \frac{ \langle \psi_{R \uparrow, k_1+k}^\dagger \psi_{R \uparrow, k_1 + k} - \psi_{R \downarrow, k_1}^\dagger \psi_{R \downarrow, k_1}\rangle}{\omega + v k - B - \frac{1}{2} gM + i0}, \nonumber\\
\chi^\pm_{0,L L}(k, \omega) &=& \frac{1}{N} \sum_{k_1} \frac{ \langle \psi_{L \uparrow, k_1+k}^\dagger \psi_{L \uparrow, k_1 + k} - \psi_{L \downarrow, k_1}^\dagger \psi_{L \downarrow, k_1}\rangle}{\omega - v k - B - \frac{1}{2} gM + i0}.
\label{eq:21} 
\eea
Observing that the denominator does not depend on the summation momentum $k_1$, we evaluate the numerator at zero temperature as 
\bea
&&\frac{1}{N} \sum_{p} \langle \psi_{R \uparrow, p+k}^\dagger \psi_{R \uparrow, p + k} - \psi_{R \downarrow, p}^\dagger \psi_{R \downarrow, p}\rangle = \int_{-\infty}^{k_{R,F,\uparrow} - k} \frac{d p}{2\pi} - \int_{-\infty}^{k_{R,F,\downarrow}} \frac{d p}{2\pi} = \frac{k_{R,F,\uparrow} -k_{R,F,\downarrow} - k}{2\pi} = \frac{B + gM/2 - v k}{2\pi v} \nonumber\\
&&= \chi_0(B + gM/2 - vk) = M - \chi_0 v k,
\label{eq:22}
\eea
where we used that dispersion of spin-$s$ right movers is $\epsilon_{R,s}(p) = v p - s(B/2 + gM/4)$, so that $k_{R,F,s} = sB_{\rm tot}/(2v)$ and
$k_{R,F,\uparrow} -k_{R,F,\downarrow} = B_{\rm tot}/v = (B + gM/2)/v$, and used \eqref{eq:9}. This leads to
\be
\chi^\pm_{0,R R}(k, \omega) =  \frac{M - \chi_0 v k}{\omega + v k - B - \frac{1}{2} gM + i0}, 
~\chi^\pm_{0,L L}(k, \omega) = \frac{M + \chi_0 v k}{\omega - v k - B - \frac{1}{2} gM + i0}.
\label{eq:23} 
\ee
This result can be checked by considering the limit of vanishing magnetic filed $B\to 0$, where we recover the expected result for the uniform susceptibility of non-interacting spinon gas \cite{maslov2015}
\be
\chi_0^\pm(k, \omega) = \chi^\pm_{0, R R}(k, \omega) + \chi^\pm_{0, L L}(k, \omega) = 2\chi_0 \frac{v^2 k^2}{\omega^2 - v^2 k^2}.
\label{eq:24}
\ee
Straightforward algebra gives 
\be
\chi^\pm(k, \omega) =  \frac{M (\omega -B - gM) + \chi_0 (1 + g\chi_0/2) v^2 k^2}{\sqrt{\tilde{v}^2 k^2 + g^2 M^2/4}} \Big(\frac{1}{\omega - \omega_{+}(k)} - 
\frac{1}{\omega - \omega_{-}(k)}\Big)
\label{eq:30}
\ee
where we introduced renormalized velocity 
\be
\tilde{v} = v \sqrt{1-g^2\chi_0^2/4}.
\label{eq:25}
\ee
Next we use \eqref{eq:10} to write $\chi_0 (1 + g\chi_0/2) v^2 = \chi \tilde{v}^2$ and observe that
\be
\frac{\omega}{\omega - \omega_{+}(k)} - \frac{\omega}{\omega - \omega_{-}(k)} = \frac{\omega_{+}(k)}{\omega - \omega_{+}(k)} - \frac{\omega_{-}(k)}{\omega - \omega_{-}(k)} .
\ee
Denoting $\Delta = g M/2$ and simplifying \eqref{eq:30} can be brought into the final form, Eq.\eqref{eq:chi_pm}, quoted in the main text 
\begin{eqnarray}
\label{eq:31}
\chi^\pm(k,\omega) & = &  M \left( \frac{A_+(k)}{\omega-\omega_+(k)}+\frac{A_-(k)}{\omega-\omega_-(k)}\right) ,\\
A_{\pm}(k) & = & 1 \pm \frac{ \tilde{v}^2 k^2 - B\Delta }{  B\sqrt{\Delta^2 + \tilde{v}^2 k^2} }, \nonumber\\
\omega_{\pm}(k) & = &  B + \Delta \pm \sqrt{\Delta^2+ \tilde{v}^2 k^2} . 
\label{eq:32}
\end{eqnarray}
Note the crucial feature of  $k^2$ scaling of the residue $A_{+}(k)$ of the upward branch $\omega_{+}(k)$ for $k\to 0$. More generally, we find
\bea
A_\pm(k) \approx 1 \mp \big(1 - \frac{(B+2\Delta)\tilde{v}^2 k^2}{2 B \Delta^2} \big)
\label{eq:33}
\eea
in the $k\to 0$ limit. Another nice feature \eqref{eq:31} shows up in the vicinity of the momentum $k_1$ at which the downward branch $\omega_{-}$ touches zero,  $\omega_{-}(k_1) =0$.
It is easy to find that for $k \approx k_1$
\be
A_{-}(k) \approx \Big(\frac{B+2\Delta}{B+\Delta}\Big)^2 \big(1-\frac{k}{k_1}\big),
\label{eq:34}
\ee
so that the residue of $\omega_{-}$ branch vanishes linearly near the touching point. Note that $k \leq k_1$ and $A_{-}$ remains positive for all $k \in (0,k_1)$.
Note that this branch does not extend beyond $k_1$. Eq.\eqref{eq:34} too is in an excellent agreement with our numerical data, see 
Figure \ref{fig:DynamicalCorrelationsSmallB}(a,c) for which $k_1 \approx \pi/4$.

\section{Two-magnon bound states and spin dynamics in the high magnetization regime}

\subsection{Bound states}

We consider the isotropic case, $\eta=1$.  The Hamiltonian  is
\begin{equation}
H = \frac{1}{2} \sum_{\ell = \pm 1, \pm 2} J_\ell \sum_n \vec{S}_n \cdot \vec{S}_{n+\ell} - B \sum_n S^z_n
\end{equation}
The ground state $|0\rangle$, of energy $E_0$, is fully polarized for strong enough $B$. %Here $J_1 = 1$ and $J_2 = \beta$, and $s=1/2$.

Let $|1k\rangle = \sum_m e^{i k m} \hat{S}_m^{-} |0\rangle$ be the 1-magnon state with momentum $k$.
It satisfies $H |1k\rangle = E_1(k) |1k\rangle$ which can be re-written as 
\begin{equation} 
(E_1(k) - E_0) |1k\rangle =\sum_m e^{i k m} [H, \hat{S}_m^{-}] |0\rangle.
\end{equation}
Carrying out the commutator and projecting onto $|0\rangle$ we of course obtain single magnon dispersion
$\epsilon_1(k) = E_1(k) - E_0 = J_1 (\cos k -1) + J_2 (\cos 2k -1) + B$. It is easy to check that for $0 < J_2 \leq J_1/4$ its minimum is at $k=\pi$, and
$\epsilon_1(\pi) = B - 2J_1$, independent of $J_2$.

The 2-magnon state is constructed as $|2 \rangle = \sum_{i,j} \psi_{ij} \hat{S}_i^{-} \hat{S}_j^{-} |0\rangle$, and its Schrodinger equation reads
\begin{equation}
(E_2 - E_0) |2\rangle =\sum_{ij}\psi_{ij} [H, \hat{S}_i^{-} \hat{S}_j^{-}] |0\rangle.
\end{equation}
Long algebra gives, using $S^z_n |0\rangle = \frac{1}{2}|0\rangle$,
\begin{eqnarray}
&&(E_2 - E_0 - 2B) \sum_{i,j} \psi_{ij} \hat{S}_i^{-} \hat{S}_j^{-} |0\rangle = \frac{1}{2}\sum_\ell J_\ell \sum_{ij} \psi_{ij} \Big( -2 \hat{S}_i^{-} \hat{S}_j^{-}
+ \hat{S}_i^{-} \hat{S}_j^{-} (\delta_{i+\ell,j} + \delta_{j+\ell,i} )\nonumber\\
&&+ \hat{S}_i^{-} \hat{S}_{j+\ell}^{-} ++ \hat{S}_{i+\ell}^{-} \hat{S}_{j+\ell}^{-} - 2 \delta_{i,j} \hat{S}_{i+\ell}^{-} \hat{S}_{i}^{-}\Big) |0\rangle .
\end{eqnarray}
Next, act on both sides of this with $\langle 0| \hat{S}_n^{+} \hat{S}_{m}^{+}$. The LHS reduces to $2(1-\delta_{n,m})\psi_{nm}$.
The RHS reads
\begin{eqnarray}
&&-4(J_1+J_2) \psi_{nm} + 2 J_{n-m} \psi_{nm} + \frac{1}{2}\sum_\ell J_\ell\Big(\psi_{n,m-\ell} + \psi_{m,n-\ell} + \psi_{n-\ell,m} + \psi_{m-\ell,n}\Big)\nonumber\\
&&-\sum_\ell J_\ell (\psi_{m,m} \delta_{n,m+\ell} + \psi_{n,n} \delta_{m,n+\ell}).
\label{eq1}
\end{eqnarray}
Note that by construction $\psi_{n,m}=\psi_{m,n}$ and that $\psi_{m,m}$ is {\em not} defined -- the wave function $|2 \rangle$ simply does not contain
the $i=j$ term since $(\hat{S}_i^{-})^2 =0$. 

Now introduce the conserved momentum $K$ of the pair and the relative momentum $q$ via
\begin{equation}
\psi_{n,m} = A e^{i K (n+m)/2} F(n-m) = A e^{i K (n+m)/2} \sum_q e^{i q (n-m)} f(q).
\label{eq2}
\end{equation}

Substitute \eqref{eq2} into \eqref{eq1}
and obtain difference equation for $F(n) = F(|n|)$. Doing so it is convenient to exclude zero relative distance when both spin flips take place
at the same site. For $s=1/2$ such a term is not present in the wave function $|2\rangle$. We therefore require $F(0)=0$.

We then obtain
\begin{eqnarray}
(E_2 - E_0 - 2 h + 2(J_1 + J_2)) F(n-m) &=& \sum'_{\ell = 1,2} J_\ell \cos[K \ell/2] (F(n-m+\ell) + F(n-m-\ell)) + \nonumber\\
&& + J_{n-m} F(n-m).
\label{eq3}
\end{eqnarray}
The prime on the sum here means summation over the positive $\ell = 1, 2$ only.

In the case of $J_1 - J_2$ chain this equation holds for $n-m \geq 3$, while for smaller relative distance (namely, 1 and 2) one needs
to modify the equation and exclude $F(0)$ terms that appear there. Then, denoting $F(n) = F_n$ and $\epsilon = E_2 - E_0 - 2 B + 2(J_1 + J_2)$, we obtain
\begin{eqnarray}
\label{eq5a}
&(\epsilon - J_1) F_1 = \xi_1 F_2 + \xi_2 (F_3 + F_1),\qquad & \text{for}~ n-m = 1\\
\label{eq5b}
&(\epsilon - J_2) F_2 = \xi_1 (F_3 + F_1) + \xi_2 F_4, \qquad & \text{for}~ n-m=2\\
&\epsilon F_n = \xi_1 (F_{n+1} + F_{n-1}) + \xi_2 (F_{n+2} +
   F_{n-2}), \qquad & \text{for}~ n \geq 3.
\end{eqnarray}
Here $\xi_1 = J_1 \cos[K/2]$ and $\xi_2 = J_2 \cos[K]$.

Now we turn the last equation into the transfer matrix form by writing
\begin{equation}
F_{n+2} = - F_{n-2} + \frac{\epsilon}{\xi_2} F_n - \frac{\xi_1}{\xi_2} (F_{n+1} + F_{n-1}).
\end{equation}
Even better, form a column vector of length 4 and then
\begin{eqnarray}
\left( \begin{array}{c}
F_{n+2} \\
F_{n+1} \\
F_n\\
F_{n-1} \end{array} \right)
=
\left( \begin{array}{cccc}
- \frac{\xi_1}{\xi_2} & \frac{\epsilon}{\xi_2} & - \frac{\xi_1}{\xi_2} & -1 \\
1 & 0 & 0 & 0 \\
0 & 1 & 0 & 0 \\
0 & 0 & 1 & 0 \end{array} \right) 
\left( \begin{array}{c}
F_{n+1} \\
F_{n} \\
F_{n-1}\\
F_{n-2} \end{array} \right) = 
M 
\left( \begin{array}{c}
F_{n+1} \\
F_{n} \\
F_{n-1}\\
F_{n-2} \end{array} \right) 
\end{eqnarray}
Eigenproblem of $M$ is solved by writing 
\begin{eqnarray} 
M \left( \begin{array}{c}
f_4 \\
f_3 \\
f_2\\
f_1 \end{array} \right) = 
\lambda \left( \begin{array}{c}
f_4 \\
f_3 \\
f_2\\
f_1 \end{array} \right) 
\end{eqnarray}
and realizing that it gives $f_2 = \lambda f_1, f_3 = \lambda f_2 = \lambda^2 f_1, f_4 = \lambda f_3 = \lambda^3 f_1$ so that the top most
line of the eigenvalue problem turns into equation for the eigenvalue
\begin{equation}
\lambda^4 + 1 + \frac{\xi_1}{\xi_2} \lambda^2 (\lambda + \frac{1}{\lambda}) - \frac{\epsilon}{\xi_2} \lambda^2 = 0
\label{eq:eigenvalue}
\end{equation}
Dividing by $\lambda^2$ one easily derives quadratic equation for $y = \lambda + \frac{1}{\lambda}$. In terms of $y$ the actual eigenvalue
is given by $\lambda = (y \pm \sqrt{y^2 - 4})/2$, where the relative sign is to be chosen so that to describe the {\em localized} solution 
for which $|\lambda| < 1$. The (un-normalized) eigenvector corresponding to the eigenvalue $\lambda$ is simply
\begin{eqnarray}
\psi \equiv \left( \begin{array}{c}
\psi_4 \\
\psi_3 \\
\psi_2 \\
\psi_1 \end{array} \right) = 
\left( \begin{array}{c}
\lambda^3 \\
\lambda^2 \\
\lambda \\
1 \end{array} \right) .
\label{eq:psi}
\end{eqnarray}
In our case there are two eigenvalues, $\lambda_1 = e^{-\kappa_1}$ and $\lambda_2 = e^{-\kappa_2}$. Hence
the solution of the transfer matrix problem, for $n\geq 1$ is given by 
\begin{eqnarray}
\left( \begin{array}{c}
F_{n+4} \\
F_{n+3} \\
F_{n+2} \\
F_{n+1} \end{array} \right) = a_1 e^{-\kappa_1 n} \psi_1 + a_2 e^{-\kappa_2 n} \psi_2
\end{eqnarray}
where $\psi_{1,2}$ are eigenvectors corresponding to $\lambda_{1,2}$.

The initial vector $(F_4, F_3, F_2, F_1)^T$ is determined by ``irregular" equations \eqref{eq5a} and \eqref{eq5b}.
These can be brought into a matrix form too
\begin{eqnarray}
\label{os11}
\left( \begin{array}{cc}
(\epsilon - J_1 -\xi_2) \psi_{11} - \xi_1 \psi_{12} - \xi_2 \psi_{13}; & (\epsilon - J_1 -\xi_2) \psi_{21} - \xi_1 \psi_{22} - \xi_2 \psi_{23} \\
(\epsilon - J_2) \psi_{12} - \xi_1 (\psi_{13} + \psi_{11})  - \xi_2 \psi_{14}; &  (\epsilon - J_2) \psi_{22} - \xi_1 (\psi_{23} + \psi_{21})  - \xi_2 \psi_{24} 
\end{array} \right)
\left( \begin{array}{c}
a_1\\
a_2 \end{array} \right) = 0
\end{eqnarray}
The above equation requires that determinant of the $2\times 2$ matrix is zero. This gives implicit 
equation on the energy of the bound state $\epsilon$. The components $\psi_{n=1,2; j = 4,3,2,1}$ of the eigenvectors are given, via equation 
\eqref{eq:psi}, by the powers of the corresponding eigenvalue $\lambda_n$. This equation can in general be solved numerically. 
Below it is applied to two important cases which are easy to treat analytically.

\subsubsection{Nearest neighbor chain: $J_2 =0$.}

We set $\xi = J_1 \cos[K/2]$ and obtain
\begin{equation}
(\epsilon - J_1) F_1 = \xi F_2, \epsilon F_2 = \xi (F_1 + F_3),  \epsilon F_3 = \xi (F_2 + F_4), ...
\end{equation}
Let $F_n = F_1 e^{-\gamma (n-1)}$ for $n\geq 2$ and obtain
\begin{equation}
\epsilon - J_1 = \xi e^{-\gamma}, \epsilon = \xi (e^{-\gamma} + e^{\gamma}).
\end{equation}
Therefore $e^{-\gamma} = \xi/J_1 = \cos[K/2]$ and $\epsilon = E_2 - E_0 - 2 B + 2 J_1 = J_1 (1 + \cos^2[K/2])$. The 2-magnon bound state with momentum $K$
has dispersion
\begin{equation}
\epsilon_2(K) = 2B - J_1 \sin^2[K/2] .
\end{equation}
Apparently this result is due to Bethe \cite{Bethe1931}.

We can also normalize the bound state wavefunction by requiring $\langle 2|2\rangle = 1$, which gives 
\begin{equation}
\sum_{i,j} \sum_{n,m} \psi^*_{nm} \psi_{ij} \langle0| \hat{S}_m^{+}\hat{S}_n^{+} \hat{S}_i^{-} \hat{S}_j^{-} |0\rangle = 2 \sum_{n,m} |\psi_{nm}|^2 = 4 N A^2 F_1^2 \sum_{r=1}^\infty e^{-2\gamma (r-1)} = \frac{N (2 A F_1)^2}{\sin^2[K/2]}.
\end{equation}
where we used that by \eqref{eq2} $|\psi_{nm}| = |A F(n-m)|$ and therefore $2 \sum_{n,m} |\psi_{nm}|^2 = 4 N A^2 \sum_{r>0} F_r^2$. 
Thus the bound state wavefunction reads
\begin{equation}
\psi_{nm} = \frac{1}{2\sqrt{N}} |\sin[K/2]| e^{i K(n+m)/2} e^{-\gamma (|n-m|-1)}.
\label{os4}
\end{equation} 

\subsubsection{$J_1-J_2$ chain: 2-magnon pair with center of mass momentum $K=\pi$.}

Here $\xi_1 = 0, \xi_2 = J_2 \cos[K] = -J_2$. This leads to the following chain of equations
\begin{equation}
(\epsilon-J_1) F_1 = \xi_2 (F_1 + F_3), (\epsilon-J_2) F_2=\xi_2 F_4, \epsilon F_n = \xi_2 (F_{n-2} + F_{n+2}) ~{\rm for} ~n\geq 3
\end{equation}
Importantly, equations for $F_n$ with $n={\rm even}$ and $n={\rm odd}$ decouple from each other, due to $\xi_1 =0$. 
Let $F_n = A e^{-\gamma n}$ and look on the {\em odd} sequence:
\begin{equation}
\epsilon-J_1 = -J_2 (1 + e^{-2\gamma}), \epsilon = -J_2 (e^{2\gamma} + e^{-2\gamma})
\end{equation}
which leads to $e^{2\gamma} = (J_2 - J_1)/J_2 < 0$ for $J_1 > 0$ and $J_2 \in (0, 1/4)$. Therefore we set $\gamma = i \pi/2 + \kappa$ and find
$e^\kappa = \sqrt{(J_1 - J_2)/J_2}$. Then
\begin{equation}
F_{n={\rm odd}} = A (-i)^n \big(\frac{J_2}{J_1 - J_2}\big)^{n/2}, \epsilon = J_1 - J_2 + \frac{J_2^2}{J_1 - J_2}.
\end{equation}
It is easy to check that for the {\em even} sequence one finds $F_{n={\rm even}} = A (-i)^n$ which does not represent a localized solution! So that only 
$F_{n={\rm odd}}$ is present. Its energy, at $K=\pi$,
\begin{equation}
\epsilon_2(\pi) = E_2 - E_0 = 2B - 2(J_1 + J_2) + \epsilon = 2B - J_1 -3 J_2 + \frac{J_2^2}{J_1 - J_2}.
\label{os10}
\end{equation}
This result has previously been obtained in \cite{Chubukov1991,Kecke2007}.

\subsection{Spin dynamics near saturation}

We now investigate how bound states appear in the dynamical spin susceptibility of the spin chain near the saturation magnetization, $1 - 2 M \ll 1$.
The structure factor
\begin{equation}
S^{+-}(q,\omega) = \int_{-\infty}^\infty dt e^{i \omega t} \sum_x e^{-i q x} \langle \psi| S^+(x,t) S^-(0,0)| \psi\rangle = 
\int_{-\infty}^\infty dt e^{i \omega t} \frac{1}{N}\sum_{m, \ell} e^{i q (m -\ell)} \langle \psi| S^+_m(t) S^-_{\ell}|\psi\rangle.
\label{os1}
\end{equation}
The average is over the ground state $|\psi\rangle$ of the chain near the saturation. We imagine a state with very small density of down spins and consider here 
state with just {\em one} down spin, $|\psi\rangle \to |\psi_Q\rangle = N^{-1/2} \sum_n e^{i Q n} S^-_n |0\rangle$. This is just 1-magnon state with momentum $Q$.
Therefore the matrix element in \eqref{os1} is proportional to $\langle \psi_Q | S^+_m(t) |A\rangle$, where 
$|A\rangle = N^{-1} \sum_{n,\ell} e^{i Q n - i q \ell} S^{-}_\ell S^{-}_n |0\rangle$. We can expand this state as 
\be
|A\rangle = \langle 2|A\rangle |2\rangle + {\rm (scattering ~states)}
\ee
The overlap with the 2-magnon bound state is found as 
\be
\langle 2|A\rangle = \frac{1}{N} \sum_{i,j} \sum_{n,\ell} \psi^*_{ij}  e^{i Q n - i q \ell} \langle0| \hat{S}_j^{+}\hat{S}_i^{+} \hat{S}_\ell^{-} \hat{S}_n^{-} |0\rangle 
= \frac{2}{N} \sum_{n,\ell} \psi^*_{\ell n}  e^{i Q n - i q \ell}
\ee
And therefore 
\be
\langle \psi_Q | S^+_m(t) |A\rangle \sim \langle \psi_Q | e^{i H t} S^+_m e^{- i H t} |2\rangle = \langle \psi_Q | S^+_m |2\rangle  e^{i \epsilon_1(Q) t} 
e^{- i \epsilon_2(K) t},
\ee
where $K$ is the momentum of the 2-magnon bound state. Finally,
\be
\frac{1}{\sqrt{N}} e^{i q m} \langle \psi_Q| S^+_m |2\rangle = \frac{1}{N} \sum_{n,m} e^{- i Q n + i q m} \langle 0|S^+_n S^+_m |2\rangle = 
\frac{2}{N} \sum_{n,m} e^{- i Q n + i q m} \psi_{nm}.
\ee
Combining these equations together we obtain
\bea
S^{+-}(q,\omega) &=& \int_{-\infty}^\infty dt e^{i \omega t} e^{i \epsilon_1(Q) t - i \epsilon_2(K) t} \frac{1}{\sqrt{N}} e^{i q m} \langle \psi_Q| S^+_m |2\rangle 
\langle 2|A\rangle \nonumber\\
&=& 2\pi \delta(\omega + \epsilon_1(Q) - \epsilon_2(K)]) |R(Q,q)|^2
\label{os6}
\eea
where, using \eqref{os4} (and therefore specializing to the nearest neighbor Heisenberg chain with $J_2=0$), 
\bea
&&R(Q,q) = \frac{2}{N} \sum_{n,m} e^{- i Q n + i q m} \psi_{nm} = \frac{1}{N^{3/2}} \sum_{n,m} e^{- i Q n + i q m} |\sin[K/2]| e^{i K (n+m)/2} e^{-\gamma(|n-m|-1)} \\
&&= \frac{1}{\sqrt{N}} \delta_{K, Q-q}  |\sin[K/2]| \sum_{r\neq 0} e^{- i (Q + q) r/2} e^{-\gamma(|n-m|-1)} = \frac{2}{\sqrt{N}} \delta_{K, Q-q}  |\sin[K/2]| \frac{\cos[(Q+q)/2] - e^{-\gamma}}{1 + e^{-2\gamma} - 2 e^{-\gamma} \cos[(Q+q)/2]}. \nonumber
\label{os5}
\eea
Using $e^{-\gamma} = \cos[K/2]$ and some trig identities we obtain
\be
R(Q,q) = - \frac{8}{\sqrt{N}} \delta_{K, Q-q}  |\sin[(Q-q)/2]| \frac{\sin[Q/2] \sin[q/2]}{3 - 2 \cos[Q] + \cos[Q-q] - 2\cos[q]} .
\label{os7}
\ee
Energy of state $|\psi_Q\rangle$ is minimal for $Q=\pi$ (for $0 \leq J_2 \leq J_1/4$) and therefore we set $Q=\pi$. Then 
\be
|R(\pi,q)|^2 = \frac{16 \delta_{K,\pi-q}}{N} \frac{\sin^2[q]}{5-3\cos[q]}
\ee
and we obtain that \eqref{os6} is peaked at $\omega = \epsilon_2(\pi-q) - \epsilon_1(\pi) = B + J_1 (3-\cos[q])/2$,
\be
S^{+-}(q,\omega) = \frac{32 \pi}{N} ~\frac{\sin^2[q]}{5-3\cos[q]} \delta(\omega - B - \frac{J_1}{2} (3-\cos[q])) .
\label{os8}
\ee
Notice that in agreement with general Larmor theorem arguments as well as with numerical results, the residue of the collective mode vanishes in the $q\to 0$ limit.
Eq.\eqref{os8} describes dispersing composite excitation with momentum $q$ which consists of a magnon at momentum $\pi$ and a 2-magnon bound 
state with momentum $\pi-q$. Within our approximation $B \to B_{\rm sat} = 2J_1$ and therefore $S^{+-}(q,\omega)$ describes mode dispersing upward from $\omega = 3 J_1$ at $q=0$ to $\omega = 4J_1$ at $q=\pi$, see Figure \ref{fig:DynamicalCorrelationsLargeB}(b). 

Notice also that the spectral weight of the 2-magnon bound state comes with $1/N$ prefactor in \eqref{os8}. This is because $S^{+-}(q,\omega)$ is calculated in the state with just {\em one} down spin, i.e. in the state with magnetization (per site) $M = 1/2 - 1/N$. Thus $1/N = 1/2 - M$, as explained in the main text, and more generally $S^{+-}(q,\omega)$ in \eqref{os8} is proportional to $1/2 - M$ in the dilute limit near the saturation, i.e. as long as $1/2 - M \ll 1$. 

The arguments presented here make it clear that there is nothing special about $J_2=0$. And indeed, taking a look at the general $J_1-J_2$ chain and {\em focusing} on $q=0$ (which corresponds to the 2-magnon state with momentum $\pi$), we observe that $S^{+-}(0,\omega)$ is peaked at (here we use \eqref{os10})
$\omega = \epsilon_2(\pi) - \epsilon_1(\pi) = 2B - J_1 -3 J_2 + \frac{J_2^2}{J_1 - J_2} - (B - 2J_1) = B + J_1 -3 J_2 + \frac{J_2^2}{J_1 - J_2} \to 3(J_1 - J_2) + \frac{J_2^2}{J_1 - J_2}$. In particular, for $J_2=J_1/4$ we observe that the bound-state peak is at $\omega = 7J_1/3$. Importantly, it is separated from the Zeeman mode, whose energy at $q=0$ is $h_{\rm sat} = 2J_1$, by a ``gap" of magnitude $J_1/3$, in a complete agreement with our numerical results, see Figure \ref{fig:DynamicalCorrelationsLargeB}(d). Away from $q=0$ point $S^{+-}(q,\omega)$ is obtained numerically, by solving \eqref{os11} for the bound state energy.

\section{Transverse correlations in the non-interacting limit}

In this section we discuss the dynamical correlations in the non-interacting limit of the Hamiltonian~\eqref{eq:H_fermions} in the main text, i.e. for $\eta=0$.
Consider the transverse correlation function, Eq.~\eqref{eq:Spm}, in the spectral representation
\begin{equation}\label{eq_sm:Spm_spectral}
S^{+-}(k, \omega)=\sum_m \left| \left\langle m\right| S^{-}_k \left|0\right\rangle \right|^2\delta(\omega-E_m)
\end{equation}
In the non-interacting limit, the ground state $\left|0\right\rangle$ and all excited states $\left|m\right\rangle$ are Slater-determinant states.
More specifically, the ground state is a filled Fermi sea with the Fermi momenta set by the magnetization $k_{{\rm F},1(2)}=\pi (1 \pm 2M)$. (Below we denote by $k_{\rm F}$ the shift in the Fermi momenta with respect to $\pi$, i. e. $2\pi M$.)
The matrix element above is given by
\begin{equation}
\left\langle m\right| S^{-}_k \left|0\right\rangle = \left\langle m\right| \sum_x e^{ikx} S^{-}_x \left|0\right\rangle = \left\langle m\right| \sum_x e^{ikx} \prod_{y<x}(-1)^{n_y} c^\dagger_x \left|0\right\rangle
\end{equation}

We now focus on the set of excited states which are single particle excited states, i.e. $\left|k'\right\rangle \equiv c^\dagger_{k'}\left|0\right\rangle=\sum_{x'}e^{ik'x'}c^\dagger_{x'}\left|0\right\rangle$ with $k' <k_{{\rm F},1}$ or $k' > k_{{\rm F},2}$. As we show below these states form the dominant contribution to the correlations function leading to the splitting of the single-magnon band as observed in Fig.~\ref{fig:DynamicalCorrelationsLargeB} of the main text.
The matrix element in this case is given by 
\begin{equation}
\left\langle k' \right| S^{-}_k \left|0\right\rangle = \sum_{x,x'} e^{i(kx-k'x')} \left\langle 0\right|  c_{x'}\prod_{y<x}(-1)^{n_y} c^\dagger_x \left|0\right\rangle
\end{equation}
The latter expectation value can be expressed as a Pfaffian of a submatrix of the covariance matrix, which describes the two-point correlations in the ground state $\left|0\right\rangle$~\cite{Bravyi2017}, and can be calculated numerically. To this end, we consider states with an even number $N_\downarrow$ of down spins and diagonalize the single particle Hamiltonian (Eq.~\eqref{eq:H_fermions} of the main text with $\eta=0$) on a finite chain of length $N=200$ with periodic boundary conditions. To make sure the single-particle eigenstates correspond to a well-defined momentum we thread a tiny flux $\sim10^{-5}$ through the closed chain. Given the eigenstates, the covariance matrix corresponding to a filled Fermi sea with $N_\downarrow$ fermions can be obtained and the matrix element above can be calculated following Ref.~\cite{Bravyi2017}.

\begin{figure}
    \centering
    \begin{overpic}[width=0.35\textwidth]{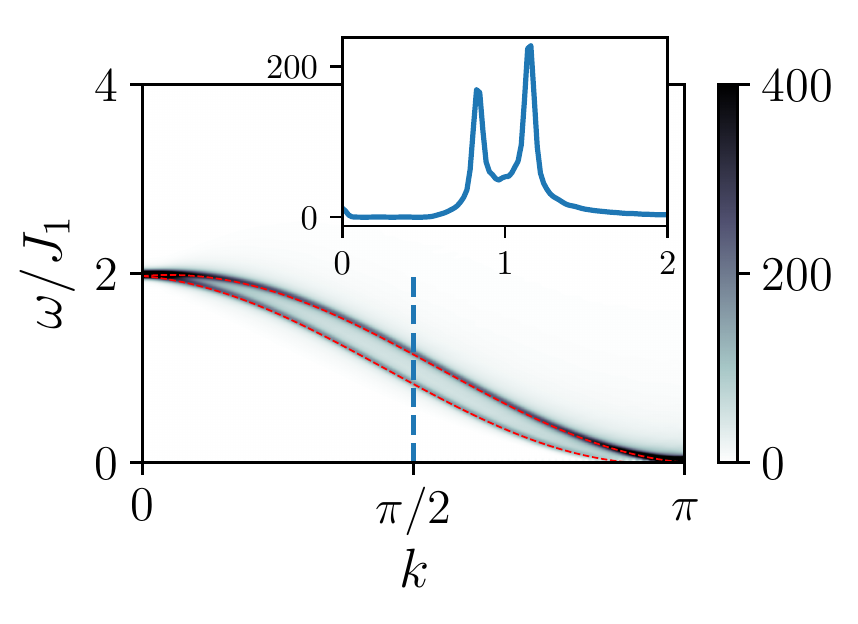} \put (0,70) {\footnotesize{(a)}} \end{overpic}
    \begin{overpic}[width=0.35\textwidth]{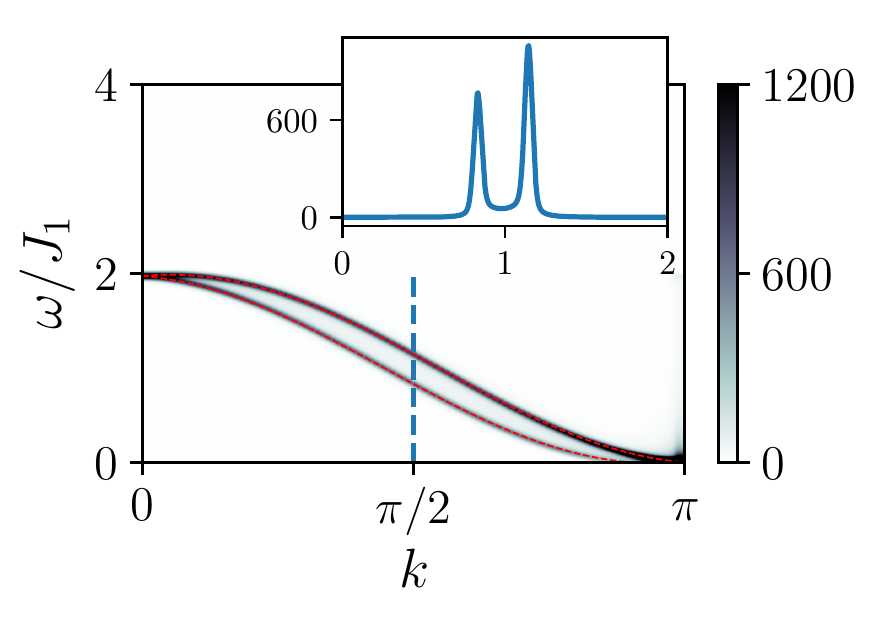} \put (0,70) {\footnotesize{(b)}} \end{overpic}
\caption{Transverse dynamical correlations for $J_2=0$ and $\eta=0$ obtained using DMRG in and time-evolution in (a) and using the exact eigenstates and calculating the matrix elements involving the Jordan-Wigner string numerically in (b) for magnetization of $M=0.45$. The insets show a cut of the response function at a fixed $k=\pi/2$ as indicated by the blue dashed line. The red dashed lines correspond to the single magnon dispersion shifted by $\pm k_{\rm F}$.}
\label{fig_sm:DMRG_vs_Noninteracting}
\end{figure}

In Fig.~\ref{fig_sm:DMRG_vs_Noninteracting} we compare the correlation function obtained used DMRG and time-evolution as in the main text, to the one obtained using the spectral representation in Eq.~\eqref{eq_sm:Spm_spectral} considering the set of exact eigenstates ${\left|k'\right\rangle}$ and calculating the matrix elements as outlined above, for a magnetization of $M=1/2-N_\downarrow/N=0.45$. 
%for different values of the magnetization $M=1/2-N_\downarrow/L$. 
As can be seen, the main features of the response function, namely, the splitting of the single magnon band by $\pm k_{\rm F}$, as well as the relative intensities of the two branches, are indeed captured in the latter approach. It is also apparent that in the DMRG calculation there is additional spectral weight within the split band. This weight is presumably due to excited states hosting additional particle-hole excitations, which were not included in the calculation above but can also be accounted for numerically.

\end{document}